\documentclass[twocolumn,amsfonts,showpacs,superscriptaddress]{revtex4-1}
\usepackage[utf8]{inputenc}

\usepackage{amssymb,amsmath,amsthm,booktabs,mathtools}
\usepackage{bbm}
\usepackage{bm}
\usepackage{color}
\usepackage{hyperref}
\usepackage{tikz}
\usepackage{pgfplots}
\usepackage{enumitem}
\usepackage{esint}
\usepackage[normalem]{ulem}

\newcommand{\comI}[1]{{\color{blue}#1}}

\date{September 2020}

\begin{document}
\title{Breakdown of Tan's relation in  lossy one-dimensional Bose gases}

\date{\today}
\author{I. Bouchoule}
\affiliation{Laboratoire Charles Fabry, Institut d’Optique, CNRS, Universit\'e Paris Sud 11, 2 Avenue Augustin Fresnel, 91127 Palaiseau Cedex, France}
\author{J. Dubail }
\affiliation{Universit\'e de Lorraine, CNRS, LPCT, F-54000 Nancy, France}

\begin{abstract}
  In quantum gases with contact repulsion, the distribution of momenta of the atoms typically decays as $\sim 1/|p|^4$ at large momentum $p$. Tan's relation connects the amplitude of that $1/|p|^4$ tail to the adiabatic derivative of the energy with respect to the gas' coupling constant or scattering length. Here it is shown that the relation breaks down in the one-dimensional Bose gas with contact repulsion, for a peculiar class of stationary states. These states exist thanks to the infinite number of conserved quantities in the system, and they are characterized by a rapidity distribution which itself decreases as $1/|p|^4$. In the momentum distribution, that rapidity tail adds to the usual Tan contact term. Remarkably, atom losses, which are ubiquitous in experiments, do produce such peculiar states. The development of the tail of the rapidity distribution originates from the ghost singularity of the wavefunction immediately after each loss event. This phenomenon is discussed for arbitrary interaction strengths, and it is supported by exact calculations in the two asymptotic regimes of infinite and weak repulsion.
\end{abstract}

\maketitle

\paragraph*{Introduction.}
In a quantum gas, contact interactions can impart large momenta to the particles: the singularity of the many-body wavefunction when two particles are at the same position is reflected in the tails of their momentum distribution $w(p)$, which decay as $w(p) \sim 1/|p|^4$. It contrasts with the gaussian decay that would be expected from the Boltzmann distribution in an ideal gas. The $1/|p|^4$ tails were noticed in hard-core one-dimensional (1D) bosons by Minguzzi {\it et al}~\cite{minguzzi_high-momentum_2002} (see also Ref.~\cite{rigol2004universal}), 
then studied in 1D gases of arbitrary interaction strength by Olshanii and Dunjko~\cite{olshanii_short-distance_2003}, 
and by Tan in  three-dimensional (3D)
fermionic gases~\cite{tan2008energetics,tan_large_2008,tan2008generalized}. [For a general analysis in two and three dimensions for bosons, fermions and mixtures, see Refs.~\cite{werner_general_2012-1,werner_general_2012}.]
Remarkably, the amplitude of the tail, $C := \lim_{p\rightarrow\infty} |p|^4 w(p)$, is a thermodynamic quantity~\cite{olshanii_short-distance_2003,tan_large_2008}.  Tan's `adiabatic sweep theorem'~\cite{tan_large_2008}, or simply `Tan's relation', connects the amplitude $C$ to the adiabatic derivative~\footnote{The derivative acts on each eigenstate, which is a function of $g$.} of the energy with respect to the two-body interaction parameter. For Bose gases, Tan's relation reads~\cite{werner_general_2012}
\begin{equation}
C = C_{\rm c} , \quad {\rm with} \quad C_{\rm c} := \frac{m^2}{(2\pi \hbar)^d}\, 2 g^2 \frac{\partial (E/V)}{\partial g} .
\label{eq:Tansintro}
\end{equation}
Here $m$ is the particles' mass, $E$ is the energy of the gas, $V$ is its volume, and $g$ is the interaction coupling constant~\footnote{In 3D, $g=4\pi\hbar^2a/m$, where $a$ is the scattering length. In 2D, $g=-2\pi \hbar^2/m/\ln(\pi a/l)$, where $l$ is a cuttoff which fulfills $l \gg a$ and $a$ is the 2D scattering length.}. The momentum distribution is normalized as $\int d^{d}p \, w(p)=N/V$, where $N$ is the total atom
number and $d$ is the dimension of the system. 
The contact density $C_{\rm c}$ is defined by the second equality of Eq.~\eqref{eq:Tansintro}, for any density matrix diagonal in the eigenbasis. Tan's relation $C=C_{\rm c}$ has been proved with wide generality and applies to many states of the gas~\cite{braaten2008exact,barth2011tan}.


Tails in the momentum distribution have been observed experimentally in 3D fermionic gases and Tan's relation has been verified~\cite{kuhnle2010universal,stewart2010verification}. It has also been verified, using spectroscopy, in 3D Bose gases~\cite{wild_measurements_2012}. On the theory side, Tan's relation and its extensions have been thoroughly investigated~\cite{braaten2008exact,zhang2009universal,braaten2010short,barth2011tan,werner_general_2012-1,werner_general_2012,rakhimov_tans_2020}. Recent works have focused on the 1D Bose gas~\cite{vignolo_universal_2013,xu2015universal,yao_tans_2018,rizzi2018scaling}, exploiting the relation between the contact density and the zero-distance two-body correlation function (Eq.~(\ref{eq:contactg2}) below).

Tan's relation (\ref{eq:Tansintro}) is based on the assumption that the tails of the momentum distribution are entirely due to the contact two-body interaction. In this Letter, we point out that this assumption is not always valid. We show that, owing to its integrability, the 1D Bose gas with contact interactions can have a contribution to its $1/|p|^4$ tail of different origin, so that $C > C_{\rm c}$. This happens in a peculiar class of stationary states, which we characterize.

Importantly, such peculiar stationary states are generated by atom losses. That makes them ubiquitous in modern cold atoms experiments in 1D~\cite{bloch_many-body_2008,bouchoule_1d_2011}, which always suffer from losses~\cite{sesko1989collisional,soding1999three,tolra2004observation}. We stress that those states are stationary with respect to Hamiltonian dynamics, so
even if losses are no longer present at long times, the breakdown
of Tan's relation persists. Therefore, an important implication of our findings is that Tan's relation will most probably be violated experimentally in 1D Bose gases.

The essence of the breakdown of Tan's relation for a gas submitted to losses is as follows.
Immediately after a loss event, the wavefunction has a singularity at the position of the lost atoms, in addition to the singularities when two of the remaining particles meet. In the momentum distribution, this additional singularity  is reflected as a $1/p^4$ term which adds to the usual contact term.
If the gas were chaotic, then it would relax to a new thermal equilibrium state. The effect on the momentum distribution would therefore be observable only at short time after the loss, since thermal states belong to the class of states that fulfill Tan's relation. However, the 1D Bose gas is not chaotic and the effect remains present even after relaxation to a stationary state.

The results presented in this Letter are twofold. First, we characterize the class of states for which Tan's relation is violated,
and we provide a formula that supersedes it (Eq.~(\ref{eq:csommecccr}) below). Second, we demonstrate that
losses bring the gas to such a state.  Our results on losses are supported by exact analyses in the hard-core and quasicondensate regimes, for which we can exploit recent results of Refs.~\cite{bouchoule_effect_2020,bouchoule_cooling_2018,johnson_long-lived_2017}. In both regimes, we find that the amplitude of the tail of the momentum distribution $C$ becomes substantially larger than the value $C_{\rm r}$ predicted by Tan's relation.

\vspace{0.1cm}
\paragraph*{The contact in the 1D Bose gas.}
We consider bosons with contact repulsion
in a periodic system of size $L$. The Hamiltonian is (with $[\Psi(z), \Psi^+(z')]=\delta(z-z')$)
\begin{equation}
  H=\int_0^L dz \, \Psi^+(z) \left( -\frac{\hbar^2 \partial_z^2}{2m} + \frac{g}{2} \Psi^+ (z)\Psi(z) \right) \Psi(z).
  \label{eq:H}
\end{equation}
We start by recalling the effects of the contact interaction on the tails of the momentum
distribution, following Ref.~\cite{olshanii_short-distance_2003}. Because of the contact interaction, the many-body wavefunction $\psi(z_1, \dots, z_N) = \left< 0 \right| \Psi(z_1) \dots \Psi(z_N) \left| \psi \right>$ has a cusp singularity whenever two positions coincide~\footnote{The cusp relation is verified by any eigenstates, and any superposition state whose eigenstate decomposition is not too irregular.}: $\partial_{z_i} \psi_{|_{z_i\rightarrow z_j^-}}-\partial_{z_i} \psi_{|_{z_i\rightarrow z_j^+}}
=(mg/\hbar^2)\,\psi(\dots,z_i=z_j,\dots)$.
When one takes the Fourier transform, those cusps become $1/p^2$ tails, which give a $\sim 1/p^4$ contribution
to the momentum distribution after taking the squared modulus of the wavefunction. When this calculation is done carefully (as in Ref.~\cite{olshanii_short-distance_2003}), it shows that the contact interaction contributes to the tail of the momentum
distribution $w(p)$ as $C_{\rm c}/p^4$ with
\begin{equation}
  \label{eq:contactg2}
C_{\rm c}= \frac{m^2}{2 \pi \hbar} g^2 n^2  g^{(2)}(0) .
\end{equation} 
Here $n=N/L$ is the atom density and $g^{(j)}(0)=\langle \Psi(z)^{+j}  \Psi(z)^j \rangle /n^j$, where $j\in {\mathbb N}$, is the normalized zero-distance $j$-body correlation function, independent of $z$ in a translation invariant system. 
Eq.~(\ref{eq:contactg2}) is an alternative, more general, definition of the contact density $C_{\rm c}$ in 1D, which works for all states including non-stationary ones. For stationary states (diagonal density matrices), it is equivalent to the one in Eq. (\ref{eq:Tansintro}). Indeed, if $\left| \psi \right>$ is an eigenstate, a straightforward application of the Hellmann-Feynman theorem leads to $n^2 g^{(2)}(0) = 2 \left< \psi \right| \partial H / \partial g  \left| \psi \right>/L = 2 \partial (E/L)/ \partial g$. 

We now argue that there exist peculiar states, not considered in Ref.~\cite{olshanii_short-distance_2003}, where the equality $C=C_c$ breaks down.

\vspace{0.1cm}
\paragraph*{The rapidity distribution, its tails, and tails of the momentum distribution.} Because of the extensive number of its conserved quantities, the 1D Bose gas typically relaxes to a Generalized Gibbs Ensemble~(see e.g. the volume \cite{calabrese2016introduction}) which is parametrised by its rapidity distribution~\cite{caux2012constructing,mossel2012generalized,ilievski2016string}. 
The rapidities are conserved by the Hamiltonian dynamics: they characterize the eigenstates of the Hamiltonian (\ref{eq:H}), which take the form of Bethe states~\cite{lieb1963exact,korepin1997quantum}. The rapidities are the asymptotic momenta of the atoms if one lets the gas expand freely in 1D~\cite{jukic2008free,bolech2012long,campbell2015sudden,caux2019hydrodynamics,wilson2020observation}. They are conveniently thought of as the momenta of  quasiparticles with infinite lifetime~\cite{bertini2016transport,bulchandani2018bethe}, dubbed `Bethe quasiparticles' in this Letter. 
After relaxation to a Generalized Gibbs Ensemble, expectation values of local observables are functionals of the rapidity distribution $\rho(q)$~\cite{caux2012constructing,mossel2012generalized,ilievski2016string}. In the following, we normalize the rapidity distribution as $\int dq \, \rho(q)= N/L$.

We stress that the rapidity distribution is {\it not equal}
to the momentum distribution of the atoms. This is well illustrated by the ground state of the system: its rapidity distribution $\rho(k)$ vanishes outside a finite interval~\cite{lieb1963exact,korepin1997quantum}, while its momentum distribution $w(p)$ presents the aforementioned $1/p^4$ tails that extend to infinity~\cite{olshanii_short-distance_2003}.


Nevertheless, for large rapidities the momentum distribution may reflect features of the rapidity distribution, and vice-versa.
To be more precise, let us imagine that the rapidity distribution of the gas has tails decaying as $1/q^4$ (we will argue below that atom losses naturally produce such tails), and let $C_{\rm r} \, := \, \lim_{q\rightarrow \infty} q^4\rho(q)$ 
be their amplitude. Then we argue below that
\begin{equation}
  C := \lim_{p \rightarrow \infty} p^4 w(p) = C_{\rm c} + C_{\rm r} .
  \label{eq:csommecccr}
\end{equation}
This formula, which generalizes Eq.~\eqref{eq:Tansintro}, is our first main result. In states where $C_r=0$, which include single eigenstates of $H$ in finite size,  thermal states, states produced by merging two thermal clouds with different temperatures~\cite{de2018edge}, Tan's relation (\ref{eq:Tansintro}) is recovered. 
On the other hand,
a non-vanishing $C_{\rm r}$ results in its breakdown. We note that Eq.~(\ref{eq:csommecccr}) can also be applied to non-stationary ones~\footnote{as long as the cusp singularity condition
  $\partial_{z_i} \psi_{|_{z_i\rightarrow z_j^-}}-\partial_{z_i} \psi_{|_{z_i\rightarrow z_j^+}}
  =(mg/\hbar^2)\,\psi(\dots,z_i=z_j,\dots)$ is fulfilled} if one uses Eq.~(\ref{eq:contactg2}) to define $C_{\rm c}$.

\vspace{0.1cm}
\paragraph*{Derivation of Eq.~(\ref{eq:csommecccr}).} We develop separate arguments for the hard-core regime $g\rightarrow \infty$ and for finite $g$. When $g\rightarrow \infty$, exact formulas are available~\cite{lenard1964momentum,vaidya1979one,SM} for the correlation function $g^{(1)}(z) = \left< \Psi^{+}(z) \Psi(0) \right>/n$, which allow us to infer its short-distance behavior. For a rapidity distribution $\rho(q)$ with a $C_{\rm r}/q^4$ tail, we find~\cite{SM}
\begin{equation}
    \label{eq:g1_shortz}
g^{(1)}(z) \underset{z \rightarrow 0} = 1 - i \frac{q_1}{n} z - \frac{q_2}{n} z^2 + i \frac{q_3}{n} z^3 + \frac{\pi (C_{\rm r} + C_{\rm c})}{6 \hbar^3 n}  |z|^3 + O(z^4),    
\end{equation}
where $q_j = \frac{1}{\hbar^j j!} \int q^j \rho(q) dq$. We arrive at this result by studying a lattice regularization of the Bose gas, for which we use an exact finite-distance formula for the two-point correlation function, and then by taking the continuum limit~\cite{SM}. Eq.~(\ref{eq:g1_shortz}) generalizes known formulas for the short-$z$ expansion of $g^{(1)}(z)$ in the $g\rightarrow \infty$ limit~\cite{olshanii_short-distance_2003,vaidya1979one,jimbo1980density} to the case of arbitrary rapidity distributions, including those with a $C_{\rm r}/q^4$ tail. 
We then use the fact that the Fourier transform of a cusp singularity in $|z|^j$ has tails decaying as $1/p^{j+1}$. Evaluating that Fourier transform, we obtain $w(p) = \frac{n}{2\pi \hbar} \int e^{i p z/\hbar} g^{(1)}(z) dz \underset{|p| \rightarrow \infty }{\simeq} (C_{\rm r} + C_{\rm p})/p^4$. Thus we arrive at Eq.~(\ref{eq:csommecccr}).

For finite $g$ and arbitrary rapidity distributions, a direct computation of the momentum distribution or of its Fourier transform $g^{(1)}(z)$ is much more difficult, even numerically~(see e.g. Refs.~\cite{Caux_2007,caux2019hydrodynamics}). Instead, we turn to a different argument, which formalizes the physical intuition that Bethe quasiparticles with large rapidities q must correspond to atoms with large
momenta $p\simeq q$. We give a brief sketch of the argument here, in order to convey the main physical idea. Details are deferred to the Supplemental Material~\cite{SM}.

Let us introduce a cutoff $\Lambda$, large enough so that $\rho(q) \simeq C_{\rm r}/q^4$ as soon as $q > \Lambda$. We split the rapidity distribution into two terms $\rho_{<\Lambda}(q)=\theta(\Lambda^2-q^2)\rho(k)$
and $\rho_{>\Lambda}(q)=\theta(q^2-\Lambda^2)\rho(q)$, where $\theta(.)$ is the Heaviside step function.
Then one can think of the gas as a two-component fluid. The 
idea is to 
take $\Lambda$ large enough so that $\Lambda \gg   {\rm max}[\frac{m g}{\hbar} ,( \frac{C_{\rm r} \xi m g}{\hbar}  )^{1/4}]$, where $\xi$ is the correlation length of the gas.

We focus first on the component with rapidity distribution $\rho_{>\Lambda}$. Within a cell of size $\gtrsim \xi$,~ large enough so that the particles it contains are not correlated with the rest of the system 
the typical number of rapidities in an interval $[q,q+dq]$ is $\xi \rho_{> \Lambda}(q) dq$. This implies that the typical spacing between neighbour rapidities is of order $\Delta q \sim 1/(\xi \rho_{>\Lambda}) \sim 1/(\xi C_{\rm r}/\Lambda^4) \gg m g/\hbar$. This ensures that this fluid component behaves as an ideal Bose gas.
In particular, its momentum distribution equals its rapidity distribution: $w_{>\Lambda}(p) \simeq \rho_{>\Lambda }(p) \simeq \theta(p^2-\Lambda^2) C_r/p^4 $.
Moreover the condition $\Lambda \gg {\rm max}[\frac{m g}{\hbar} ,( \frac{C_{\rm r} \xi m g}{\hbar}  )^{1/4}]$ also ensures that the two fluid components do not interact between each other.

The other fluid component is characterized by a rapidity distribution $\rho_{< \Lambda}$ with no tails, so it satisfies Tan's relation. Thus, its momentum distribution $w_{< \Lambda} (p)$ decays as $C_{\rm c}/p^4$ at large $p$.

The total momentum distribution $w(p)$ of the gas is the sum of the momentum distributions of both components, which leads to Eq.~(\ref{eq:csommecccr}).

Having established the key formula~(\ref{eq:csommecccr}), we now turn to the question: Is there a physical process that produces such peculiar states with $1/q^4$ tails in their rapidity distribution? We are aware of only one such example in the literature so far: a sudden quench of the interaction strength $g$, which relaxes to a state with $C_{\rm r} >0$~\cite{de2014solution}. In the rest of this Letter we argue that atom losses, which are ubiquitous in experiments, always generate these peculiar states.

\vspace{0.1cm}
\paragraph*{Losses and $1/q^4$ tails of the rapidity distribution.} We consider the general case of local $K$-body losses, where $K=1,2,3, \dots$ is the number of atoms lost in each loss event. Depending on the experiment, losses are typically dominated by $K=1$,  $K=2$~\cite{hensler2003dipolar,kinoshita2005local} or $K=3$ processes~\cite{soding1999three,tolra2004observation}, but it is convenient to keep $K$ arbitrary. The atom density then decays as $dn/dt=- K G g^{(K)}(0) n^K$, where
$G$ is a constant with units of ${\rm length}^{K-1}.{\rm time}^{-1}$ that characterizes the loss rate.
Following Ref.~\cite{bouchoule_effect_2020} (see also Refs.~\cite{lange2018time,rossini2020strong}), we assume that the loss rate $G n^{K-1}$ is much smaller than the relaxation time, so that the gas relaxes to a Generalized Gibbs Ensemble after each loss event. This allows to represent the evolution of the gas under losses by its time-dependent rapidity distribution~\cite{bouchoule_effect_2020}.

Let us assume that, at $t=0$ the gas' rapidity distribution has no $1/q^4$ tails, {\it i.e.}  
$C_{\rm r}(t=0) = 0$. For instance, the gas could be in a thermal state. We want to show that at, $t=0$, $d C_{\rm r}/dt > 0$, implying that the rapidity distribution will develop non-vanishing $1/q^4$ tails.

To do this, we elaborate on the microscopic mechanism presented in the introduction. Consider the many-body wavefunction $\psi_{t=t_l^-}(z_1,\dots ,z_N)$ just before a loss event occuring at time $t_l$ and position $z_l$. Right after the loss, the wavefunction of the remaining $N-K$ atoms is
$
\tilde{\psi}_{t=t_l^+}(z_1,\dots,z_{N-K})=L^{K/2} 
\psi_{t=t_l^-}(z_1,\dots,z_{N-K},z_{N-K+1}=z_l,\dots,z_{N}=z_l).
$
As a reminiscence of its cusp singularities before the loss, the wavefunction $\tilde\psi_{t= t_l^+}$ still has a cusp  at $z_j = z_l$ ($j=1,\dots,N-K$).
Following the calculation of Ref.~\cite{olshanii_short-distance_2003}, we find
that it results in a contribution $C^{(1\, {\rm loss})}/p^4$ to the momentum distribution, with the amplitude
 \begin{equation}
   \begin{array}{rcl}
  C^{(1\, {\rm loss})}&=& \frac{\hbar^3}{2\pi} {L^{K-1}(N-K)} \int dz_2 \dots dz_{N-K}  \\
   && \qquad \quad | \partial_{z_1} \psi_{|_{z_1\rightarrow z_l^+}}- 
   \partial_{z_1} \psi_{|_{z_1\rightarrow z_l^-}}|^2,
   \end{array}
   \label{eq:c1lderivee}
 \end{equation}
 where the variables $z_{N-K+1},\dots ,z_N$ in the integrand
 are taken equal to $z_l$. The boundary condition
imposed by the contact interaction gives  
$ \partial_{z_1} \psi_{|_{z_1\rightarrow z_l^-}}-
 \partial_{z_1} \psi_{|_{z_1\rightarrow z_l^+}}= 
K mg/\hbar^2  \, \psi(z_1=z_l,z_2, \dots ,z_{N-K+1}=z_l,
\dots z_N=z_l)$. 
Then, using the expression
of $g^{(K+1)}(0)$ in first quantization, we get
\begin{equation}
         C^{(1 \, {\rm loss})}    =  \frac{m^2}{2\pi \hbar} \,\frac{n K^2}{L } g^2g^{(K+1)}(0) .
 \label{eq:c1lg2}
\end{equation}
Here we have used the fact that, as $N\rightarrow\infty$,
$N-K\simeq N$ and $N\dots (N-K)\simeq N^{K+1}$.

Next, we rely on formula (\ref{eq:csommecccr}), and argue that the contribution (\ref{eq:c1lg2}) of one loss event to the momentum distribution translates into the {\it same} contribution to the rapidity distribution. Indeed, the contribution (\ref{eq:c1lg2}) is not taken into account 
in the contact density $C_{\rm c}$ at time $t= t_l^+$, therefore according to formula (\ref{eq:csommecccr}) it must appear in the tail of the rapidity distribution:
\begin{equation}
    C_{{\rm r} |_{t = t_l^+}} - C_{{\rm r} |_{t = t_l^-}} = C^{(1 \, {\rm loss})}.
\end{equation}
Like $\rho(k)$, $C_r$ is conserved by the Hamiltonian dynamics, so this increase of $C_{\rm r}$ remains after relaxation to a Generalized Gibbs Ensemble. Finally, we multiply this result by $L G n^Kg^{(K)}(0) dt$, the number of loss events occuring in the system during a short time interval $dt$. This leads to the initial growth rate
 \begin{equation}
    \frac{dC_{\rm r}}{dt} (t=0) \,= \, \frac{m^2}{2\pi \hbar} {Gn^{K+1}}K^2g^2 g^{(K)}(0)g^{(K+1)}(0) .
   \label{eq:dcrdti}
 \end{equation}
 This equation is the second main formula of this Letter. 
 It shows that $dC_{\rm r}/dt_{|_{t=0}}>0$, such that $C_{\rm r}$ becomes non-zero. Together with Eq.~(\ref{eq:csommecccr}), it implies that the momentum distribution develops tails that are larger than what is expected from Tan's relation.

We stress that Eq.~\eqref{eq:dcrdti} gives only the initial  growth rate of the tail of the rapidity distribution. At later times, its evolution will also involve additional damping effects. Indeed, under atom losses the gas ultimately evolves to the vacuum, therefore the whole rapidity distribution ---including its tails--- will go to zero at very long times. The calculation of the damping of $C_{\rm r}$ at longer times is not obvious. Below we obtain further results in the hard-core and quasicondensate regimes.

\vspace{0.1cm}

\paragraph*{Exact results in the hard-core regime.} In the hard-core regime ($g \rightarrow \infty$), only one-body losses are relevant, since $g^{(K)}(0)=0$ for $K>1$. 
Thus, in this paragraph we fix $K=1$.
The evolution of the rapidity distribution $\rho(t,q)$
 under losses has been computed recently in 
 Ref.~\cite{bouchoule_effect_2020}, for an arbitrary initial distribution $\rho(t=0,q)$, see in particular Eq.~(14) in that reference. Here we exploit that general result to study the evolution of the $1/q^4$ tail.

Expanding Eq.~(14) of Ref.~\cite{bouchoule_effect_2020} for large $q$, we find that $\rho(t,q) = C_{\rm r}(t)/q^4 + {\rm o}(1/q^4)$, with
\begin{equation}
    C_{\rm r}(t)= \frac{4\hbar m}{\pi} \, [n(0)e(0) - j(0)^2/(2m)] \,e^{-G t}(1-e^{-G t}).
    \label{eq:Cr_time}
\end{equation}
Here $j(t) = \int q \, \rho(t,q)  dq$ and $e(t) = \int q^2/(2m) \, \rho(t,q)  dq$ are the momentum and energy density respectively~\footnote{$n_0e_0 - j_0^2/(2m) =n_0 e_i$, where $e_i$ is the energy density of the gas in the reference frame where the center of mass is at rest.}.
The right hand side of Eq.~(\ref{eq:Cr_time}) involves these quantities at time $t=0$. Using the fact that, under losses, the particle, momentum, and energy densities evolve as $n(t) = n(0) e^{- G t}$, $j(t) = j(0) e^{- G t}$, $e(t) = e(0) e^{- G t}$ respectively in the $g \rightarrow \infty$ limit~\cite{SM}, the right hand side can also be written as $\frac{4\hbar m}{\pi} \, [n(t)e(t) - j(t)^2/(2m)] (e^{G t}-1)$.

We note that formula~(\ref{eq:Cr_time}) provides a non-trivial check of our general prediction (\ref{eq:dcrdti}) for the initial growth rate{\comI : using the standard identity $\lim_{g \rightarrow \infty} n^2 g^2 g^{(2)}(0) \, =\, \frac{8 \hbar^2}{m} \,[ ne - j^2/(2m)]$~\cite{SM}, one sees that Eqs.~(\ref{eq:dcrdti}) and (\ref{eq:Cr_time}) agree}.

Importantly, Eq.~(\ref{eq:Cr_time}) also allows us to compare the amplitude $C_{\rm r}(t)$ with the contact density at time $t$.  Using again the standard identity above, together with Eq.~(\ref{eq:contactg2}), we find
\begin{equation}
    C_{\rm r}(t) / C_{\rm c}(t) = \exp (G t)-1.
    \label{eq:Cr_Cc_time}
\end{equation}
We see that the ratio of the amplitude $C_{\rm r}$ to the contact density $C_{\rm c}$ grows exponentially as time increases. This is our third main result: not only does the term $C_{\rm r}/p^4$ contribute to the momentum distribution, it can also become dominant compared to the contact term. Numerical  calculations 
of $w(p)$~\cite{SM} show
that, for an initial degenerate gas,  $w(p)\simeq (C_r+C_c)/p^4$  as soon as  $p\gtrsim 7 \hbar n_0$. 

We now investigate the ratio $C_{\rm r}(t)/C_{\rm c}(t)$ for weak repulsion.

\vspace{0.1cm}
\paragraph*{Results for the quasicondensate.}
In the quasicondensate regime, correlations between atoms are weak and $g^{(j)}(0)\simeq 1$ for all $j$. An effective description of the gas is obtained by a phase-density representation~\cite{mora_extension_2003}: in Eq.~(\ref{eq:H}), one writes the atomic field $\Psi$ as $\sqrt{n + \delta n} e^{i \theta}$  where $\theta$ and $\delta n$ are  phase and density fluctuation fields (with $\delta n,\partial\theta/\partial z \ll n$), which satisfy the commutation relation  $[\delta n(z), \theta(z')] = i \delta(z-z')$. The Bogoliubov approximation then leads to a collection of independent harmonic modes. The Hamiltonan for each mode is of the form $H_{k}= \varepsilon_k b^+_k b_k$ (up to additive constant), where $b_k^+$ ($k \in \frac{2\pi \hbar}{L} \mathbb{Z}$) is a linear combination of the Fourier modes $\delta n_k$ and $\theta_k$~\cite{mora_extension_2003,SM} and $\varepsilon_q=\sqrt{\frac{k^2}{2m}(\frac{k^2}{2m}+2gn)}$. The Bogoliubov creation/annihilation operators satisfy  $[b_k, b^+_{k'}] = \delta_{k,k'}$, and the occupation of each mode is $\alpha_k = \left< b^+_k b_k \right>$.

The effect of slow losses on the Bogoliubov mode occupations $\alpha_k$ has been analyzed in Refs.~\cite{grisins_degenerate_2016,johnson_long-lived_2017,schemmer_monte_2017,bouchoule_cooling_2018}. In Ref.~\cite{bouchoule_cooling_2018}, the effect of $K$-body losses on $\alpha_k$ was computed for small $k$. 
In Refs.~\cite{grisins_degenerate_2016,johnson_long-lived_2017}, the evolution of $\alpha_k$ was studied for any $k$, but
only $K=1$ was considered.
Combining these results, we are able to compute  $d\alpha_k/dt$
for any $K$ and $k$~\cite{SM}. The result reads
\begin{equation}
  \frac{d  \alpha_k}{dt} = K^2 G n^{K-1} \left ( -\alpha_k - \frac{1}{2} + \frac{1}{4}\left[ \frac{\varepsilon_k}{k^2/(2m)} + \frac{k^2/(2m)}{\varepsilon_k}  \right] \right ).
  \label{eq:dalphaqdt}
\end{equation}

The precise link between Bogoliubov excitations and Bethe quasiparticles is not obvious. However, it has been discussed by Lieb~\cite{lieb_exact_1963} (see also Ref.~\cite{ristivojevic2014excitation}), who identifies, for states
close to the ground state, 
the large-$k$ Bogoliubov 
excitations to Bethe quasiparticles with 
rapidities $q\simeq k$. Therefore a $C_{\rm r}/q^4$ tail in the rapidity distribution translates to Bogoliubov mode occupations decaying as $\alpha_k \simeq 2\pi \hbar ~C_{\rm r}/k^4$ for large $k$~\footnote{The factor $2\pi \hbar$ comes from the fact that the number of Bogoliubov excitation in an interval $[k,k+dk]$ is 
$\alpha(k) L dk / (2\pi \hbar)$, while the number of rapidities
is $L\rho(k)dk$.}. [We have checked~\cite{SM} that this identification $q\simeq k$, together with the known exact expression for $g^{(1)}(z)$~\cite{mora_extension_2003}, is compatible with our Eq.~(\ref{eq:csommecccr}) within the framework of Bogoliubov theory,  as it should.]

Using the large-$k$ expansion of $\varepsilon_k$ in Eq.~\eqref{eq:dalphaqdt}, we find that
the amplitude of the $1/q^4$ tails of $\rho(q)$ evolves according to 
\begin{equation}
  \frac{dC_r}{dt}= K^2Gn^{K-1} \left ( -C_r  + \frac{m^2}{2\pi \hbar} \, g^2 n^2 \right ).
  \label{eq:dcbdt}
\end{equation}
This differential equation can be easily solved~\cite{SM}, which allows us to obtain $C_{\rm r}(t)$ at all times. In particular, at long times, we find that the ratio of $C_{\rm r}(t)$ to the contact density $C_{\rm c}(t)= \frac{m^2}{2\pi \hbar} g^2 n(t)^2$  (Eq.~(\ref{eq:contactg2}), with $g^{(j)}(0)=1$) behaves as
\begin{equation}
    \label{eq:CrCc_main}
    \frac{C_{\rm r}(t)}{C_{\rm c}(t)} \, \underset{t \rightarrow \infty}{=} \, \left\{ 
        \begin{array}{ccl}
            \exp (G t)  & {\rm if} & K=1, \\ 
            2 \log (G n_0^{K-1} t)  & {\rm if} & K=2, \\ 
            K/(K-2)  & {\rm if} & K \geq 3  .
        \end{array}
    \right.
\end{equation}
This is the fourth main result of this Letter. For $K=1$, one finds the same behavior as in the hard-core regime. For $K \geq 3$, the ratio takes an asymptotic value. For instance, the ratio $C_r/C_{\rm c}$ goes to $3$ for three-body losses, so  the tail of the momentum distribution $C/p^4$ is four times larger than its value predicted by Tan's relation (\ref{eq:Tansintro}).

\vspace{0.1cm}
\paragraph*{Experimental prospects.}
An experimental test of the predictions of this paper is within reach in
current cold atom setups. There exist different ways of measuring
the momentum distribution of 1D gases~\cite{fabbri_momentum-resolved_2011,shvarchuck_bose-einstein_2002,van_amerongen_yang-yang_2008,jacqmin_momentum_2012}. Because of the small amplitude of the tails, such a measurement requires a high dynamical range,
which can be achieved for instance using metastable atoms~\cite{cayla_single-atom-resolved_2018}. Usually, gases in 
experiments are non-uniform. Within a local density approximation, our results are straightforwardly generalized to include a trapping potential~\cite{SM}.

\vspace{0.1cm}
\paragraph*{Conclusion.}
On the theory side, our results open several research lines.
First, for quantitative comparison with experiment,
one should  compute the evolution
of the rapidity tails in intermediate regimes of the 1D gas. For this, one can
in principle rely on the method presented in Ref.~\cite{bouchoule_effect_2020}, although an improvement of the numerical efficiency of that method would be required (see also the recent analytical progress in Ref.~\cite{hutsalyuk2020integrability}). Second, our results can probably be extended to integrable 1D Fermi gases~\cite{guan2013fermi}. Third, it would be
interesting to investigate the effects of losses in higher dimension. The singularity of the wavefunction at the position of the lost atoms is also expected to have a effect that remains to be elucidated.
Finally, it would be interesting to study loss processes that are not purely local or not purely Markovian. How would this impact the development of the momentum tails?

\acknowledgments{We thank B. Doyon for discussions and joint work on closely related topics. We also thank A. Minguzzi, M. Olshanii, P. Vignolo, and F. Werner for useful comments on the manuscript.
This work was supported by ANR QUADY - ANR-20-CE30-0017-02.}

\bibliography{momentum_tails,momentum_tailsIsa}

\appendix

\newpage

\begin{center}
\large{{\bf Supplementary material}}
\end{center}

This Supplemental Material contains:
\begin{itemize}
    \item {\bf App. A:} a derivation of formula (5) in the main text for the short-distance behavior of $g^{(1)}(z)$ in the hard-core limit. We also present a numerical method to calculate the momentum distribution $w(p)$ from the rapidity distribution $\rho(q)$ in the hard-core limit.

    \item {\bf App. B:} the detailed argument for Eq. (4) in the main text at finite $g$
    
    \item {\bf App. C:} the calculation of the hard-core limit of the product $g^2 g^{(2)}(0)$,
    
    \item {\bf App. D:} a derivation of the fact that, under one-body losses and in the hard-core limit, the atom density, momentum density and energy density evolve simply as $n(t) = e^{-Gt } n(0)$, $j(t) = e^{-Gt } j(0)$, $e(t) = e^{-Gt } e(0)$
     
    \item {\bf App. E:} detailed calculations in the weakly interacting regime within Bogoliubov theory: the evaluation of the momentum distribution, the effect of losses on the Bogoliubov modes, and the solution of the differential equation (13),
    
    \item {\bf App. F:} a brief discussion about the generalization of our results to non-uniform gases.  
    
\end{itemize}


\section{Momentum distribution in the hard-cord limit}

In this section we set $\hbar = m =1$.

\subsection{Conjecture about $g^{(1)}$ on the lattice}

We take a lattice gas of free fermions, with creation/annihilation operators $c_j^\dagger$, $c_j$ ($j \in \mathbb{Z}$) satisfying $\{c_j, c_{j'}^\dagger \} = \delta_{j,j'}$. We consider a translation-invariant Gaussian state characterized by the two-point function $\left< c_j^\dagger c_{j'} \right> = \left< c_{j-j'}^\dagger c_{0} \right> $.
We want to study the boson one-body density matrix, which includes a Jordan-Wigner string between the two fermion operators. For $j \geq 0$, it is defined as
\begin{equation}
    \label{eq:app_def_g1latt}
	g^{(1)}_{\rm latt.}(j) := 	\left< c_j^\dagger   \prod_{a=1}^{j-1} (-1)^{c^\dagger_a c_a} c_{0} \right> ,
\end{equation}
and, for $j<0$, as $g^{(1)}_{\rm latt.}(j) := g^{(1)}_{\rm latt.}(-j)^*$. We use the following exact formula which gives $g^{(1)}_{\rm latt.}(j)$ as a $j \times j$ Toeplitz determinant~\cite{vaidya1979one},
\begin{equation}
    \label{eq:toeplitz}
	 g^{(1)}_{\rm latt.}(j) = 2^{j-1} \left| \begin{array}{cccccc} G(1) & G(2) & \dots  & G(j) \\
	G(0) & G(1) &  & \vdots \\
	\vdots &  & \ddots  & G(2) \\
	G(2-j)& \dots  & G(0) & G(1)
	 \end{array} \right| ,
\end{equation}
with
\begin{equation}
	G(j) = \left\{ \begin{array}{ccc}
		\left< c_{j}^\dagger c_{0} \right>  & {\rm if } & j \neq 0 \\
		\left< c_{j}^\dagger c_{0} \right> - \frac{1}{2}  & {\rm if } & j= 0  .
	\end{array}
	\right.
\end{equation}

Let us assume that the fermion two-point function depends on a small parameter $\epsilon > 0$, such that its expansion for $\epsilon \rightarrow 0^+$ is of the form (for $j \geq 0$)
\begin{equation}
    \qquad  \left< c_{j}^\dagger c_{0} \right> \, \underset{\epsilon \rightarrow 0^+}{=} \,  a_0 \epsilon  + a_1 j \epsilon^2 + a_2 j^2 \epsilon^3  + a_3 j^3 \epsilon^4 + O( \epsilon^5 ) ,
\end{equation}
and $\left< c_{j}^\dagger c_{0} \right> := \left< c_{-j}^\dagger c_{0} \right>^*$ if $j<0$.
Here the coefficient $a_0$ is real, but  $a_1,a_2, a_3$ can be complex. For this fermion two-point function, we want to know the small-$\epsilon$ expansion of the boson one-density matrix (\ref{eq:app_def_g1latt}). Using formula (\ref{eq:toeplitz}), we have computed that expansion with Mathematica, for small values of $j$. We find
{\small
\begin{eqnarray*}
    g^{(1)}_{\rm latt.}(1) & \underset{\epsilon \rightarrow 0^+}{=}&  a_0 \epsilon + a_1  \epsilon^2 + a_2  \epsilon^3 + a_3  \epsilon^4  + O(\epsilon^5)  \\
    g^{(1)}_{\rm latt.}(2) & = &  a_0 \epsilon + 2 a_1  \epsilon^2 + 2^2 a_2 \epsilon^3 + (2^3 a_3 - 2(2 a_0 a_2 - a_1^2)) \epsilon^4 + O(\epsilon^5)  \\
    g^{(1)}_{\rm latt.}(3) & = &  a_0 \epsilon + 3 a_1  \epsilon^2 + 3^2 a_2 \epsilon^3 + (3^3 a_3 - 8(2 a_0 a_2 - a_1^2))  \epsilon^4 + O(\epsilon^5)  \\
    g^{(1)}_{\rm latt.}(4) & = &  a_0 \epsilon + 4 a_1  \epsilon^2 + 4^2 a_2 \epsilon^4 + (4^3 a_3 - 20(2 a_0 a_2 - a_1^2))  \epsilon^4 + O(\epsilon^5)  \\
    g^{(1)}_{\rm latt.}(5) & = &  a_0 \epsilon + 5 a_1  \epsilon^2 + 5^2 a_2 \epsilon^4 + (5^3 a_3 - 40(2 a_0 a_2 - a_1^2)) \epsilon^4 + O(\epsilon^5)  \\
    g^{(1)}_{\rm latt.}(6) & = &  a_0 \epsilon + 6 a_1  \epsilon^2 + 6^2 a_2 \epsilon^4 + (6^3 a_3 - 70(2 a_0 a_2 - a_1^2)) \epsilon^4 + O(\epsilon^5)  \\
    g^{(1)}_{\rm latt.}(7) & = &  a_0 \epsilon + 7 a_1  \epsilon^2 + 7^2 a_2 \epsilon^4 + (7^3 a_3 - 112 (2 a_0 a_2 - a_1^2)) \epsilon^4 + O(\epsilon^5)  \\
    g^{(1)}_{\rm latt.}(8) & = &  a_0 \epsilon + 8 a_1  \epsilon^2 + 8^2 a_2 \epsilon^4 + (8^3 a_3 - 240 (2 a_0 a_2 - a_1^2)) \epsilon^4 + O(\epsilon^5) ,
\end{eqnarray*}
}
which leads us to the obvious conjecture (for $j \geq 0$):
\begin{eqnarray}
    \label{eq:app_conj}
  g^{(1)}_{\rm latt.}(j)  &\underset{\epsilon \rightarrow 0^+}{=}& a_0 \epsilon + a_1 j \epsilon^2 + a_2 j^2 \epsilon^3 \\
 \nonumber  && + [ a_3 j^3 - \frac{j(j^2-1)}{3}(2 a_0 a_2-a_1^2) ]  \epsilon^4  +O(\epsilon^5).
\end{eqnarray}
That calculation is of combinatorial nature, and it is probably possible to prove that formula. A proof for all $j$ is not essential for our purposes though. It is sufficient to know that it holds true for a few different values of $j$. Below, we use it to infer the short-distance behavior of the one-particle density matrix of the continuous Bose gas in the hard-core limit.

\subsection{Eq.~(5) in the main text}

We consider a continuous gas of hard core bosons in a Gaussian state characterized by its rapidity distribution $\rho(q)$. Namely, if $c^\dagger(x)$, $c(x)$ are the fermion creation/annihilation operators in the continuum, we look at a Gaussian state with a translation-invariant fermion two-point function
\begin{equation}
	\left< c^\dagger(x) c(x') \right> =   \int_{-\infty}^{\infty}  e^{- i q (x-x')} \rho(q)  dq .
\end{equation}
Let us look first at the short-distance behavior of $\left< c^\dagger(x) c(0) \right>$. When $\rho(q)$ decays sufficiently fast (say, exponentially) at large $q$, it can be obtained simply by expanding the exponential in the integral,
\begin{equation}
    \label{eq:app_exp_notail}
	\left< c^\dagger(x) c(0) \right> \underset{ x \rightarrow 0}{=} q_0 - i q_1 x - q_2 x^2 + i q_3 x^3 + O(x^4).
\end{equation}
with $q_a = \int \frac{q^a}{a!} \rho(q) dq$. When $\rho(q)$ decays as a power-law, this expansion breaks down, which is reflected in the fact that the coefficients $q_a$ are infinite for $a$ large enough. From now on we assume that $\rho(q) \simeq \frac{C_{\rm r}}{q^4}$ for $q \rightarrow \pm \infty$. The correct small-$x$ expansion is then
\begin{eqnarray}
\nonumber &&	\left< c^\dagger(x) c(0) \right> \underset{x\rightarrow 0}{=}   \\ &&  q_0 - i q_1  x  -  q_2  x^2 + i q_3  x^3  + \frac{\pi C_{\rm r}}{6}  |x|^3  +  O(x^4).  \qquad 
\end{eqnarray}
Here the coefficient $q_3$ is finite because the two divergences in the integral $\int q^3/q^4 dq$ when $q \rightarrow \pm \infty$ cancel. To obtain the term $\frac{\pi C_{\rm r}}{6} |x|^3$, one can for instance write $\rho(q)$ as $(\rho(q) - \frac{C_{\rm r}}{4+q^4}) + \frac{C_{\rm r}}{4+q^4}$. The first term does not have a tail, so it has an expansion of the form (\ref{eq:app_exp_notail}), while the Fourier transform of the second term is evaluated straightforwardly and is $\frac{\pi C_{\rm r}}{4}e^{-|x|} (\cos |x| + \sin |x|) \simeq \frac{\pi C_{\rm r}}{4} (1 - x^2 + \frac{2}{3} |x|^3 + \dots )$.

Now let us turn to the boson one-particle density matrix $g^{(1)}(x)$. We regard $g^{(1)}(x)$ as the continuum limit of $g^{(1)}_{\rm latt.}(j)$ when the lattice spacing $\epsilon$ is much smaller than the inverse density of particles $1/q_0$. Namely, for $x \in \epsilon \mathbb{Z}$,
\begin{equation}
    g^{(1)}(x) \underset{  \epsilon q_0 \ll 1 }{\simeq} \frac{1}{q_0 \epsilon} \, g^{(1)}_{\rm latt.}(x/\epsilon) .
\end{equation}
This identification must hold provided that the lattice fermion two-point function corresponds to a discretization of the continuous one. For instance we can take
\begin{eqnarray}
	&& \left< c_{j}^\dagger c_{0} \right> : = 
	\epsilon \left< c^\dagger(j \epsilon) c(0) \right> \\
\nonumber	&& =  q_0 \epsilon - i q_1 j \epsilon^2 - q_2 j^2 \epsilon^3 + i q_3 j^3 \epsilon^3 + \frac{\pi C_{\rm r}}{6}  |j|^3 \epsilon^3 +  O(a^4) .
\end{eqnarray}
We are interested in the behavior of $g^{(1)}(x)$ for small $x>0$. We have two small parameters: $x$ and the lattice spacing $\epsilon$ (or, equivalently, the dimensionless $x q_0$ and $\epsilon q_0$). Let us consider a smooth function $F(\epsilon,x)$, $\epsilon>0,x>0$,  which coincides with $\frac{1}{q_0\epsilon} \, g^{(1)}_{\rm latt.}(x/\epsilon)$ for $x \in \epsilon\mathbb{N}$. Notice that $F(0,x) = g^{(1)}(x)$.   $F(\epsilon,x)$ should have a double-expansion in the two small parameters,
\begin{equation}
    F(\epsilon,x) \, = \, \sum_{l \geq 0,m \geq 0} \alpha_{l,m} \epsilon^l x^m .
\end{equation}
We can use Eq.~(\ref{eq:app_conj}), with $a_0 = q_0$, $a_1 = -i q_1$, $a_2 = -q_2$, $a_3 = i q_3 + \frac{\pi C_{\rm r}}{6}$,  to fix the first few coefficients $\alpha_{l,m}$. Indeed, for fixed $j$,
\begin{equation}
   \frac{1}{q_0 \epsilon} g^{(1)}_{\rm latt.}(j) \, = \,  F(\epsilon, j \epsilon)   \, = \, \sum_{l \geq 0,m \geq 0} \alpha_{l,m} j^m \epsilon^{l+m}  ,
\end{equation}
so when one expands both sides for small $\epsilon$, the identification of the terms of order $O(\epsilon^{l+m})$ gives
\begin{eqnarray*}
   1 &=& \alpha_{0,0}  \\
   -i \frac{q_1}{q_0} j &=& \alpha_{1,0}+ \alpha_{0,1} j \\
   -\frac{q_2}{q_0} j^2  &=& \alpha_{2,0} + \alpha_{1,1} j + \alpha_{0,2}j^2 \\
   (i \frac{q_3}{q_0} + \frac{\pi C_{\rm r}}{6q_0}) j^3 + \qquad \quad && \\
   \qquad \frac{j(j^2-1)}{3}\frac{2 q_0 q_2-q_1^2}{q_0} &=& \alpha_{3,0} + \alpha_{2,1} j + \alpha_{1,2}j^2 + \alpha_{0,3} j^3  .
\end{eqnarray*}
Since this holds for several values of $j$, we get linearly independent equations that fix all the coefficients. In particular, we find $\alpha_{0,1} = - i \frac{q_1}{q_0}$, $\alpha_{0,2} = - \frac{q_2}{q_0}$, $\alpha_{0,3} = i \frac{q_3}{q_0} + \frac{\pi}{6 q_0} [ C_{\rm r} + \frac{4}{\pi}(q_0 q_2-q_1^2/2) ]$.

The continuous one-particle density matrix $g^{(1)}(x)$ is given by $F(0,x)$, so we obtain
\begin{eqnarray}
    g^{(1)}(x) & \underset{x \rightarrow 0^+}{=} & 1- i \frac{q_1}{q_0} x  - \frac{q_2}{q_0}x^2 + i \frac{q_3}{q_0} x^3 \\
 \nonumber  && + \frac{\pi}{6 q_0} [ C_{\rm r} + \frac{4}{\pi}(q_0 q_2-q_1^2/2) ] x^3 + O(x^4).
\end{eqnarray}
Since $g^{(1)}(-x) = g^{(1)}(x)^*$, we see that we also have
\begin{eqnarray}
    g^{(1)}(x) & \underset{x \rightarrow 0^-}{=} & 1- i \frac{q_1}{q_0} x  - \frac{q_2}{q_0}x^2 + i \frac{q_3}{q_0} x^3 \\
 \nonumber  && - \frac{\pi}{6 q_0} [ C_{\rm r} + \frac{4}{\pi}(q_0 q_2-q_1^2/2) ] x^3 + O(x^4).
\end{eqnarray}
Thus, our final result for the short-distance behavior of the one-particle density matrix is
\begin{eqnarray}
    g^{(1)}(x) & \underset{x \rightarrow 0}{=} & 1- i \frac{q_1}{q_0} x  - \frac{q_2}{q_0}x^2 + i \frac{q_3}{q_0} x^3 \\
 \nonumber  && + \frac{\pi}{6 q_0} [ C_{\rm r} + \frac{4}{\pi}(q_0 q_2-q_1^2/2) ] |x|^3 + O(x^4).
\end{eqnarray}
This is our formula (5) in the main text. The coefficient $\frac{4}{\pi}(q_0 q_2-q_1^2/2)$ is the contact density $C_{\rm c}$ in the hard-core limit. This is easily shown by combining formula (3) in the main text with $\lim_{g \rightarrow \infty} q_0^2 g^2 g^{(2)}(0) = 8 [q_0 q_2 - q_1^2/2]$ (in units with $m=\hbar=1$), see the Appendix C below.

\subsection{Numerical evaluation of the momentum distribution $w(p)$ from the rapidity distribution $\rho(q)$}

We have also studied the momentum distribution numerically in the hard-core limit, by evaluating the momentum distribution $w(p)$ of hard-core bosons as a functional of their rapidity distribution $\rho(q)$. Here we explain how we implement that procedure. In this section we set $\hbar = m =1$. We exploit formulas (14)-(15) of Ref.~\cite{atas2017exact}, which gives the one-body density matrix as follows:
\begin{equation}
    \label{eq:1pdm_TG}
   \left< \Psi^\dagger(x) \Psi(y) \right>    =  \sum_{i,j=0}^\infty \varphi_i (x) \sqrt{n_i} Q_{ij}(x,y) \sqrt{n_j} \varphi_j^* (y)  ,
\end{equation}
where the $\varphi_i(x)$ $(i=0,\dots, \infty)$ are the single-particle eigenfunctions of the Schr\"odinger operator for an infinite system in an external potential, $-\hbar^2/(2m)\partial_x^2 + V(x)$, and $n_i \in [0,1]$ is the occupation of each orbital. In Ref.~\cite{atas2017exact}, it is assumed that the $n_i$ are the occupations of a Gibbs ensemble at a given temperature and chemical potential. But Eq.~(\ref{eq:1pdm_TG}) is more general, and it holds true for any occupations, corresponding to a Generalized Gibbs Ensemble. The semi-infinite matrix  $Q(x,y)$ is defined as $Q(x,y) = (P^{-1})^T \, {\rm det}\,P$, with
\begin{equation}
   P_{ij}(x,y)  = \delta_{ij} - 2 \,{\rm sign}(y-x) \sqrt{n_i n_j} \int_{x}^y \phi_i(z) \phi_j^*(z) dz .
\end{equation}
We stress that this formula is based on the mapping from hard-core bosons to free fermions, and that it works for an infinite system. In principle, it does not apply to a finite system with periodic boundary conditions. The reason is that hard-core bosons with periodic boundary conditions map to periodic/anti-periodic boundary conditions for the fermions, depending on the whether the total number of fermions is odd/even respectively. Since formula (\ref{eq:1pdm_TG}) works for arbitrary occupation numbers, the parity of the number of fermions is not fixed (unless all $n_i$ are equal to $0$ or $1$).

\begin{figure}[htb]
    \includegraphics[width=0.43\textwidth]{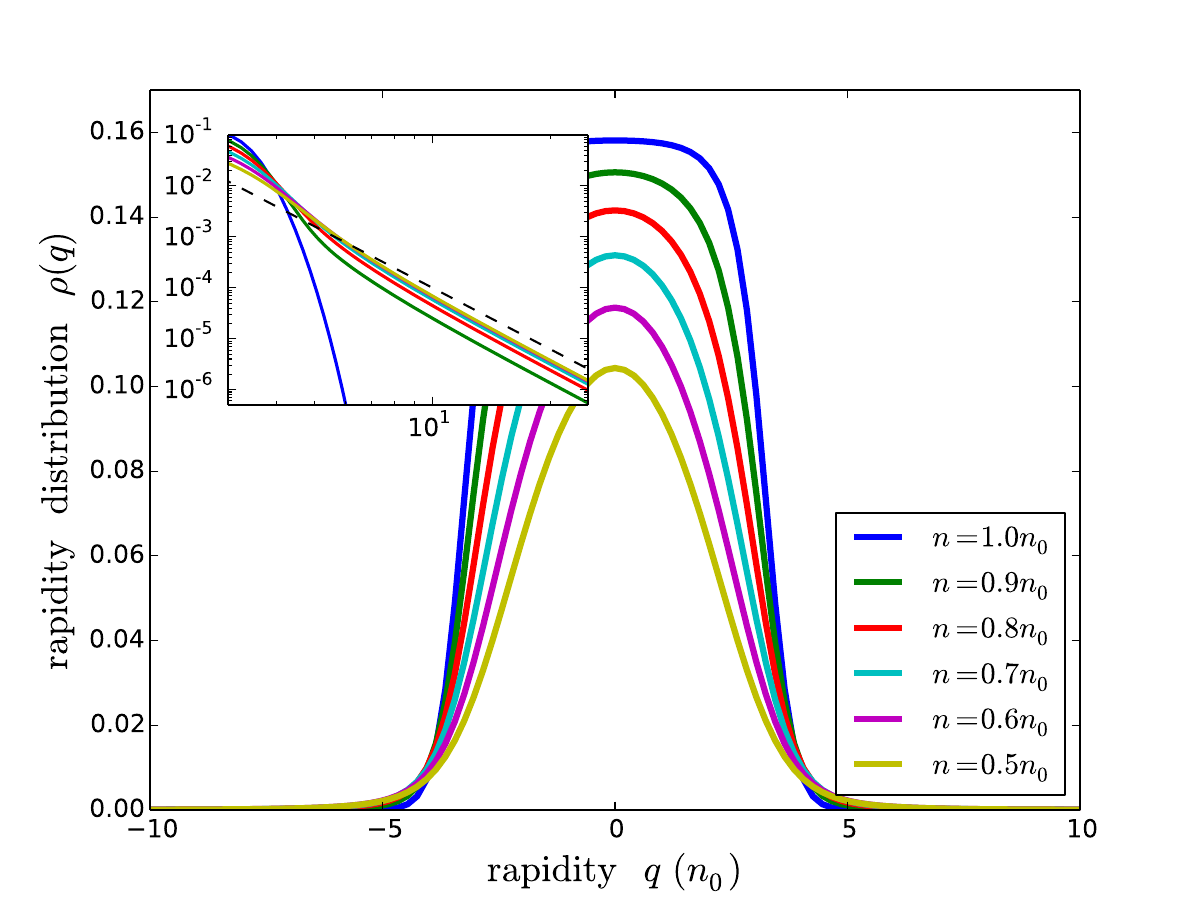} \\
    \includegraphics[width=0.43\textwidth]{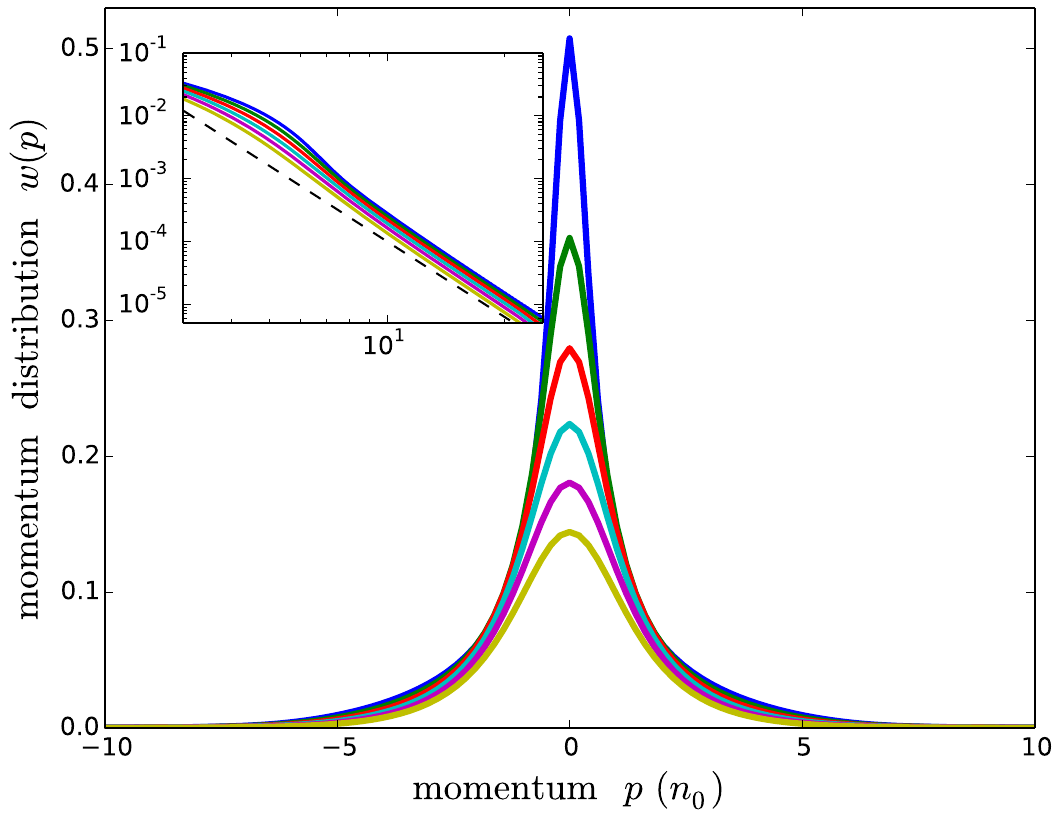}
    \caption{Top: rapidity distribution in the hard-core limit, given  by Eq.~(9) in the main text. The initial rapidity distribution $\rho_0(q)$ (blue curve) is the thermal distribution at temperature $T=1.02 n_0^2$ and chemical potential $\mu = 5T$. The other curves are the rapidity distributions after some fraction ($10\%$, $20\%$, \dots, $50\%$) of the atoms have been lost. The inset shows a zoom on the tails of $\rho(q)$ in logarithmic scale; the black dashed line is the $1/q^4$ curve. In the initial state, $\rho_0(q)$ decays as a Gaussian, but at later times $\rho(q)$ has a $\sim 1/q^4$ tail.
    Bottom: the corresponding momentum distributions, obtained from our numerical procedure. The inset shows a zoom on the tails of $w(p)$ in logarithmic scale; the black dashed line is the $1/p^4$ curve.}
    \label{fig:SM_rapidity_momentum_TG}
\end{figure}

However, the one-body density matrix typically decays quickly with the distance $|x-y|$. Moreover, we are mostly interested in its short-distance behavior, because this is what fixes the large-$p$ tail of the momentum distribution. Therefore, we can work with $x,y \in [-L/2,L/2]$ with {\it periodic boundary conditions for the fermions} as long as $L$ is large enough. Thus, we can use plane waves $\varphi_j (x) = e^{i q_j x}/\sqrt{L}$ with $q_j \in 2\pi \mathbb{Z}/L$, such that 
\begin{equation}
    \label{eq:1pdm_TG2}
   \left< \Psi^\dagger(x) \Psi(0) \right>    \underset{L \rightarrow \infty}{=}  \frac{2\pi}{L} \sum_{q_i,k_j \in \frac{2\pi}{L}\mathbb{Z}} e^{i q_i x} \sqrt{\rho(q_i) \rho(k_j)} Q_{ij}(x,0) .
\end{equation}
Here we have used the fact that the occupation of each fermionic mode is given by the rapidity density, $n_i = 2\pi \rho(q_i)$. In practice, we numerically evaluate the right hand side of Eq. (\ref{eq:1pdm_TG2}) by truncating the sum, using a finite set of orbitals $q_i \in \{ -\frac{2\pi}{L} M, \dots, -\frac{2\pi}{L} , 0, \frac{2\pi}{L} , \dots ,  \frac{2\pi}{L} M \}$ for large enough $M$.

Finally, the momentum distribution is obtained by numerically evaluating the Fourier transform
\begin{equation}
   w(p) = \frac{1}{2\pi} \int e^{i p x} \left< \Psi^\dagger(x) \Psi(0) \right> dx .
\end{equation}
With this method, we obtain the momentum distribution $w(p)$ accurately for $1/L \ll | p| < 2\pi M/L$. In Fig.~\ref{fig:SM_rapidity_momentum_TG} we show the momentum distribution obtained for rapidity distributions corresponding to Eq.~(9) in the main text, for an initial thermal distribution at temperature $T=1.02 n_0^2$ and chemical potential $\mu = 5 T$, after some fraction of the atoms have been lost ($n_0$ is the initial density of atoms). These results are obtained with $L=31/n_0$ and $M=125$, so they are accurate for $0.03 n_0 \ll |p| < 25 n_0$. This is enough to observe the $1/p^4$ tail (see the inset of Fig.~\ref{fig:SM_rapidity_momentum_TG}, bottom).

In practice, to extract the amplitude of tail $C$, we use the values of $f(p) := p^4 w(p)$ inside a window $p \in [p_{\rm min} ,p_{\rm max}]$ where $p_{\rm min}$ is large enough such that one focuses on the tail, and $p_{\rm max}$ is small enough so that we avoid the effects of the truncation of the basis of orbitals. We then fit these values with a function $C/p^4 + \alpha_1/p^5 + \alpha_2/p^6$ to extract the coefficient $C$. This gives us access to $C$, within an error bar that is typically around $\sim 4\%$.

Alternatively, the amplitude $C$ can be extracted directly from the short-distance behavior of $\left< \Psi^\dagger(x) \Psi(0) \right>$. Numerically, this is more efficient because one does not have to compute the two-point function for many values of $x$ to evaluate the Fourier transform. One needs only a few values in a small interval $[0,\varepsilon]$, where $\varepsilon$ is chosen as some fraction of the inverse density $1/n_0$ (we choose $\varepsilon = 0.25/n_0$). Then we fit these values with a polynomial of the form $\left< \Psi^\dagger(x) \Psi(0) \right> = n_0 + \alpha_2 x^2 + \frac{\pi C}{6} x^3 + \alpha_4 x^4 + \alpha_5 x^5 + \alpha_6 x^6$, which gives us access to $C$. The precision of this procedure is higher, and we obtain $C$ with an error of order $0.5 \%$. This is mainly due to the fact that, since we need to compute less points, we can use much larger numbers of orbitals in our truncated sum (\ref{eq:1pdm_TG2}). We use $\sim 6000$ orbitals (corresponding to $M \sim 3000$, compared to $M=125$ above).

We find that the amplitude $C$ obtained with this method always satisfies Eq. (4) in the main text.

\section{Detailed argument for Eq.~(4) in the main text at finite $g$}

Here we elaborate on the derivation of the formula $C= C_{\rm c} + C_{\rm r}$ sketched in the main text. The main physical intuition behind this argument is that Bethe quasi-particles with large rapidities $\lambda$ must correspond to atoms with large momenta $p \simeq \lambda$. We start by making that intuition more precise at the level of Bethe states. In this section we set $m = \hbar =1$.

\subsection{Preliminary: factorization of Bethe states}

Let $\pmb{\lambda}_N =  \{ \lambda_1, \dots , \lambda_N \}$ be a set of rapidities, with
\begin{equation}
    \lambda_1 < \dots < \lambda_N,
\end{equation}
that satisfies the Bethe equations (see below and Ref.~\cite{korepin1997quantum}). Let $\left| \pmb{\lambda}_N \right>$ be the corresponding Bethe state, whose wavefunction is~\cite{korepin1997quantum}
\begin{eqnarray}
	\label{eq:bethe}
\nonumber  && \left< 0 \right| \Psi (x_1) \dots \Psi(x_N) \left| \pmb{\lambda} \right>  \\ 
\nonumber && \,  \propto \, \sum_{\sigma \in S_N} (-1)^{|\sigma|} \prod_{1 \leq a<b \leq N}  \left( \lambda_{\sigma(b) } - \lambda_{\sigma(a) }  - i g\,  {\rm sgn}(x_b-x_a) \right) \\
&& \qquad \quad \times e^{i \sum_a x_a \lambda_{\sigma(a)}}  .
\end{eqnarray}
Now let us assume that the largest rapidity is separated from the other ones by an interval much larger than $g$,
\begin{equation}
	\label{eq:assumption}
	 | \lambda_N  - \lambda_{N-1} | \gg g .
\end{equation}
Then we argue that
\begin{equation}
	\label{eq:tensor_largep}
	\left| \pmb{\lambda}_{N} \right>  \simeq  \Psi^\dagger_{\lambda_N} \left| \pmb{\lambda}_{N-1} \right>,
\end{equation}
where $\Psi^\dagger_{p} =  \frac{1}{\sqrt{L}} \int_0^L e^{i p x} \Psi^\dagger (x) dx$ is the Fourier mode of the boson creation operator $\Psi^\dagger(x)$. This is physically clear: if one boson has very large momentum $p \simeq \lambda_N$, then its interaction with the other $N-1$ bosons is almost suppressed. So the eigenstate must be a tensor product `$\Psi^\dagger_{\lambda_N} \left|  0\right> \otimes \left| \pmb{\lambda}_{N-1} \right>$'. More formally, this is seen directly at the level of Eq. (\ref{eq:bethe}): assuming (\ref{eq:assumption}), we have
\begin{eqnarray*}
&& \left< 0 \right| \Psi (x_1) \dots \Psi(x_N) \left| \pmb{\lambda} \right> \\
&& \propto   \sum_{\sigma \in S_N}  (-1)^{|\sigma|}   ( - 1)^{N- \sigma^{-1}(N)} \\
&& \prod_{a<b, \sigma(a)\neq N, \sigma(b) \neq N} \left( \lambda_{\sigma(b) } - \lambda_{\sigma(a) }  - i g\,  {\rm sgn}(x_b-x_a) \right) e^{i \sum_a x_a \lambda_{\sigma(a)}} .
\end{eqnarray*}
We set $d = \sigma^{-1} (N)$ and $\sigma' = \sigma \circ \tau_{d N}$ where $\tau_{ij}$ is the transposition $i \leftrightarrow j$, such that $\sigma'(N) = N$. Then we can sum over $d \in \{ 1, \dots , N \}$ and $\sigma' \in S_{N-1}$ separately. After some straightforward manipulations of the indices, this gives
\begin{eqnarray*}
	&& \left< 0 \right| \Psi (x_1) \dots \Psi(x_N) \left| \pmb{\lambda} \right> \\
	&& \propto  \sum_{d=1}^N  e^{i  x_{d} \lambda_{N}}   \sum_{\sigma' \in S_{N-1}}  (-1)^{|\sigma'|}  \\
	&& \prod_{1 \leq a<b \leq N-1} \left( \lambda_{\sigma(b) } - \lambda_{\sigma(a) }  - i g\,  {\rm sgn}(x_{\tau_{ d N}(b)}-x_{\tau_{d N}(a)}) \right) \\
	&& \qquad \times e^{i \sum_{a = 1}^{N-1} x_{\tau_{dN}(a)} \lambda_{\sigma(a)}} ,
\end{eqnarray*}
so that we recognize
\begin{eqnarray}
	\label{eq:tensor_largep_1st}
	&& \left< 0 \right| \Psi (x_1) \dots \Psi(x_N) \left| \pmb{\lambda} \right> \\
\nonumber	&& \quad = \mathcal{S} \cdot  e^{i x_N \lambda_N} \, \left< 0 \right| \prod_{1 \leq j  \leq N-1} \Psi (x_j)  \left| \pmb{\lambda}_{N-1} \right> ,
\end{eqnarray}
where $ \mathcal{S}$ is the symmetrizer over all indices of an $N$-variable function, i.e. $ \mathcal{S} \cdot f( x_1, \dots, x_N)  := \frac{1}{N!} \sum_{\sigma \in S_N} f(x_{\sigma(1)}, \dots , x_{\sigma(N)})$. Eq.~(\ref{eq:tensor_largep_1st}) is nothing but the first-quantized form of Eq.~(\ref{eq:tensor_largep}). 

Moreover, under the assumption (\ref{eq:assumption}), $\lambda_N$ becomes independent from the other rapidities at the level of the Bethe equations. Namely, the $N$ equations~\cite{korepin1997quantum}
\begin{equation}
    e^{i \lambda_a L} = \prod_{1 \leq b \leq N, b\neq a} \frac{\lambda_a - \lambda_b + i g}{\lambda_a - \lambda_b - i g} , \qquad a=1, \dots , N
\end{equation}
become, assuming (\ref{eq:assumption}),
\begin{eqnarray}
    \label{eq:app_bethe_decouple}
\nonumber     e^{i \lambda_a L} &=& \prod_{1 \leq b \leq N-1, b\neq a} \frac{\lambda_a - \lambda_b + i g}{\lambda_a - \lambda_b - i g} , \qquad a=1, \dots , N-1, \\
    e^{i \lambda_N L} &=& 1 .
\end{eqnarray}

\vspace{1cm}

Clearly, if one has more rapidities that are widely separated,
\begin{equation}
	\label{eq:assumptionM}
	 | \lambda_N  - \lambda_{N-1} | , | \lambda_{N-1}  - \lambda_{N-2} | , \dots, | \lambda_{N-M+1}  - \lambda_{N-M} |   \gg g ,
\end{equation}
then one gets
\begin{equation}
	\label{eq:tensor_largepM}
	\left| \pmb{\lambda}_{N} \right>  \simeq  \Psi^\dagger_{\lambda_N} \Psi^\dagger_{\lambda_{N-1}} \dots \Psi^\dagger_{\lambda_{N-M+1}} \left| \pmb{\lambda}_{N-M} \right>,
\end{equation}
in the same sense as above. This simply follows by induction on $M$.

\subsection{Model of independent cells}

We consider the following model. We take a gas in a very large box of size $L$. We assume that it has a finite correlation length $\xi$, so that we can divide it into $m$ small independent cells containing $N^{(j)}$ particles (with a total particle number $N = \sum_{j=1}^m N^{(j)}$), and of length $\ell^{(j)}$ (of order a few times the correlation length $\xi$). We further assume that the state within each cell may be represented by a single eigenstate for a small periodic system of size $\ell^{(j)}$. The eigenstate in the $j^{\rm th}$ cell is a Bethe state with rapidities $\lambda^{(j)}_{1} < \dots < \lambda^{(j)}_{N^{(j)}}$, and the rapidity distribution in the full system is taken as the sum of the rapidities in all the cells,
\begin{equation}
	\label{eq:rho_toymodel}
	\rho(\lambda)  \, := \, \frac{1}{L} \sum_{j=1}^m \left( \sum_{a=1}^{N^{(j)}}  \delta ( \lambda - \lambda_a^{(j)} ) \right) .
\end{equation}
In the $m \rightarrow \infty$ limit (which implies $L \rightarrow \infty$ since we are working with cells of fixed size of order $\xi$), Eq.~(\ref{eq:rho_toymodel}) becomes a smooth rapidity distribution. We assume that $\rho(\lambda)$ decays as $C_{\rm r} / \lambda^4$ for large $\lambda$.

\vspace{0.5cm} Now, within the framework of this model, we derive Eq.~(4) of the main text. We start by selecting a cutoff $\Lambda$ large enough so that the following conditions are satisfied:
\begin{enumerate}
    \item $\Lambda$ is much larger than the typical width of the distribution $\rho(\lambda)$, so that for $\lambda > \Lambda$, one is really in the tail of the distribution:  $\rho(\lambda) \simeq C_{\rm r}/\lambda^4$ for any $\lambda > \Lambda$ ,
    \item $\Lambda \gg g $
    \item $\Lambda^4 \gg  \xi C_{\rm r} g$.
\end{enumerate}

For a cell $j$, let $M^{(j)}$ be the number of rapidities larger than $\Lambda$ ($M^{(j)}$ can be zero). Since the rapidities are ordered we have $\lambda^{(j)}_{N^{(j)}-M^{(j)}} < \Lambda < \lambda^{(j)}_{N^{(j)}-M^{(j)}+1}$ when $M^{(j)} >  0$. Similarly, we can define $\bar{M}^{(j)}$, the number of rapidities smaller than $-\Lambda$. Because of condition 1., $M^{(j)}$ and $\bar{M}^{(j)}$ can be estimated to be of order
\begin{equation}
    \label{eq:app_Mj}
	M^{(j)} \, = \,  \ell^{(j)} \int_{\Lambda }^\infty \rho_{> \Lambda}(\lambda) d\lambda \sim \frac{ \ell^{(j)} C_{\rm r}}{\Lambda^3} \sim \frac{ \xi C_{\rm r}  }{\Lambda^3}  .
\end{equation}
There are two cases: either this is much smaller than one, or it is larger than one, depending on whether it is condition 2. or 3. that prevails.

If $\xi C_{\rm r} < g^3$, then condition 2. is more restrictive. Condition 2. implies that $\frac{\xi C_{\rm r}}{\Lambda^3} \ll 1$. In that
case, we can assume that, in each cell $j$, $M^{(j)}$ is either zero or one. In the case when $M^{(j)}$ is one, the largest rapidity $\lambda^{(j)}_{N^{(j)}}$ is distributed with a probability $p  (\lambda)  \simeq  \frac{1}{\lambda^4} / \int_\Lambda^\infty \frac{du}{u^4}$, so its distance to all the other rapidities is typically of order $\Lambda$. Consequently, condition 2. implies
\begin{equation}
	\label{eq:larger_than_c}
	| \lambda^{(j)}_{N^{(j)}}  - \lambda^{(j)}_{N^{(j)}-1}  |  \gg   g . 
\end{equation}

If $\xi C_{\rm r} > g^3$, then condition 3. is more restrictive. Condition 3. does not put a constraint on $M^{(j)}$. [This is because it leads to $\frac{\xi C_{\rm r}}{\Lambda^3} \ll \Lambda/g$, which is automatically satisfied because $\Lambda / g$ is very large.] In that case there can be several rapidities larger than $\Lambda$ in each cell $j$. In an interval $[\lambda, \lambda + \Delta \lambda]$ (with $\lambda > \Lambda$), there are typically $\xi \rho_{> \Lambda} (\lambda)\Delta \lambda \simeq \frac{\xi C_{\rm r}}{\lambda^4}\Delta \lambda$ rapidities, so the typical spacing between two rapidities is $\sim \lambda^4/ (\xi C_{\rm r} ) > \Lambda^4/ (\xi C_{\rm r} )$. Then condition 3. implies
\begin{equation}
	\label{eq:larger_than_c}
	| \lambda^{(j)}_{N^{(j)}}  - \lambda^{(j)}_{N^{(j)}-1}  | , \dots, | \lambda^{(j)}_{N^{(j)}- M^{(j)}+1}  - \lambda^{(j)}_{N^{(j)}- M^{(j)}}  |  \gg   g . 
\end{equation}

So, in both cases, we find that the $M^{(j)}$ rapidities larger than $\Lambda$ are separated from the other rapidities by an interval that is large compared to $g$. Whenever $M^{(j)} > 1$, those $M^{(j)}$ rapidities are also well separated from one other. The same discussion applies to the $\bar{M}^{(j)}$ rapidities smaller than $-\Lambda$.

\vspace{0.5cm}

We can then apply the analysis of the previous subsection in each cell $j$. The Bethe state $ \left| \pmb{\lambda}^{(j)}_{N^{(j)}} \right> $ factorizes:
\begin{eqnarray}
	\label{eq:tensor_largepMj}
\nonumber	\left| \pmb{\lambda}^{(j)}_{N^{(j)}} \right> & \simeq &  \Psi^\dagger_{\lambda^{(j)}_{N^{(j)}}} \dots \Psi^\dagger_{\lambda^{(j)}_{N^{(j)}-M^{(j)}+1}} \\
    &&	\times \Psi^\dagger_{\lambda^{(j)}_1} \dots \Psi^\dagger_{\lambda^{(j)}_{\bar{M}^{(j)}}}  \left| \pmb{\lambda}^{(j)}_{\bar{M}^{(j)}+1,N^{(j)}-M^{(j)}} \right>, \qquad 
\end{eqnarray}
where $\left| \pmb{\lambda}^{(j)}_{\bar{M}^{(j)}+1,N^{(j)}-M^{(j)}} \right>$ is the Bethe state with rapidites $\{ \lambda^{(j)}_{\bar{M}^{(j)}+1}, \lambda^{(j)}_{\bar{M}^{(j)}+2} \dots , \lambda^{(j)}_{N^{(j)}-M^{(j)}} \}$.
The momentum distribution in the cell $j$ is then given by
\begin{eqnarray}
\nonumber	 && \left< \pmb{\lambda}_{N^{(j)}} \right|  \Psi_p^\dagger  \Psi_p   \left| \pmb{\lambda}_{N^{(j)}} \right> \, \simeq \\
\nonumber	 && \qquad  \left< \pmb{\lambda}^{(j)}_{\bar{M}^{(j)}+1,N^{(j)}-M^{(j)}} \right|  \Psi_p^\dagger \Psi_p  \left| \pmb{\lambda}^{(j)}_{\bar{M}^{(j)}+1,N^{(j)}-M^{(j)}} \right> \\
 &&  \qquad + \sum_{a=1}^{\bar{M}^{(j)}} \delta (p - \lambda^{(j)}_a)  + \sum_{a=M^{(j)}+1}^{N^{(j)}} \delta (p - \lambda^{(j)}_a) ,
\end{eqnarray}
where $\Psi^\dagger_p$ creates a boson in the cell $j$ with momentum $p$. Summing over the cells and taking the $m \rightarrow \infty$ limit, we find the total momentum distribution
\begin{eqnarray}
    \label{eq:app_wp_sum}
\nonumber &&	w(p) : =  \frac{1}{L} \sum_{j=1}^m  \left< \pmb{\lambda}_{N^{(j)}} \right|  \Psi_p^\dagger \Psi_p  \left| \pmb{\lambda}_{N^{(j)}} \right> \\
\nonumber	&& \simeq   \frac{1}{L} \sum_{j=1}^m \left< \pmb{\lambda}^{(j)}_{\bar{M}^{(j)}+1,N^{(j)}-M^{(j)}} \right|  \Psi_p^\dagger \Psi_p  \left| \pmb{\lambda}^{(j)}_{\bar{M}^{(j)}+1,N^{(j)}-M^{(j)}} \right> \\
 \nonumber   && \quad + \frac{1}{L} \sum_{j=1}^m \left( \sum_{a=1}^{\bar{M}^{(j)}} \delta (p - \lambda^{(j)}_a)  + \sum_{a=M^{(j)}+1}^{N^{(j)}} \delta (p - \lambda^{(j)}_a) \right) .\\
\end{eqnarray}
In this second term, we recognize the tail of the rapidity distribution (\ref{eq:rho_toymodel}). More precisely, we can split the distribution (\ref{eq:rho_toymodel}) into two terms $\rho_{< \Lambda}(\lambda): = \rho(\lambda) \theta(|\Lambda | - \lambda )$ and $\rho_{> \Lambda}(\lambda) := \rho(\lambda) \theta(\lambda - |\Lambda | )$, where $\theta(u) = 1$ if $u\geq 0$ and $\theta(u) = 0$ otherwise. Then the second term in Eq.~(\ref{eq:app_wp_sum}) is equal to $\rho_{> \Lambda}(\lambda) \simeq \frac{C_{\rm r}}{\lambda^4} \theta(|\lambda|-\Lambda)$. The first term in~(\ref{eq:app_wp_sum}) is the momentum distribution $w_{< \Lambda}(p)$ evaluated in the macrostate with rapiditity distribution $\rho_{< \Lambda} (\lambda)$.

Thus we arrive at
\begin{eqnarray}
    \label{eq:app_CcplusCr_0}
\nonumber w(p) & \simeq & w_{< \Lambda} (p) + \rho_{> \Lambda}(p) \\
    & \underset{|p| \rightarrow \infty}{\simeq} &   \frac{C_{{\rm c}, < \Lambda} }{p^4} +  \frac{C_{{\rm r}} }{p^4} .
\end{eqnarray}
The term $C_{{\rm c}, < \Lambda} /p^4$ comes from Tan's relation, which is valid because the rapidity distribution $\rho_{< \Lambda}(\lambda)$ does not have tails. Notice that this gives the contact density $C_{{\rm c}, < \Lambda}$ evaluated in that state, as opposed to the contact density $C_{\rm c}$ evaluated in the macrostate with the initial rapidity distribution $\rho(\lambda)$.

\vspace{0.7cm}

Finally, we show that the contact density $C_{{\rm c}, < \Lambda}$ is actually equal to $C_{{\rm c}}$. To obtain the contact density, we apply the Hellmann-Feynman theorem independently to each cell. We rely again on the factorization of the Bethe state (\ref{eq:tensor_largepMj}), and on the fact that the Bethe equations for the $M^{(j)}+\bar{M}^{(j)}$ rapidities outside $[-\Lambda,\Lambda]$ decouple, as in  Eq.~(\ref{eq:app_bethe_decouple}). The fact that the Bethe equations decouple for those rapidities implies that they no longer vary with $g$, so their derivative w.r.t $g$ vanishes.  Thus we have 
\begin{eqnarray}
\nonumber &&	\frac{\partial}{\partial g} \left< \pmb{\lambda}_{N^{(j)}} \right|  H \left| \pmb{\lambda}_{N^{(j)}} \right> \\
\nonumber	&&  \simeq \frac{\partial}{\partial g} \left( \sum_{a=1}^{\bar{M}^{(j)}} \frac{(\lambda_a^{(j)})^2}{2} + \sum_{a=M^{(j)}+1}^{N^{(j)}} \frac{(\lambda_a^{(j)})^2}{2} \right. \\
\nonumber && \qquad + \left. \vphantom{\sum_{a=1}^{\bar{M}^{(j)}} } \left< \pmb{\lambda}^{(j)}_{\bar{M}^{(j)}+1,N^{(j)}-M^{(j)}} \right|  H \left| \pmb{\lambda}^{(j)}_{\bar{M}^{(j)}+1,N^{(j)}-M^{(j)}} \right> \right)  \\
\nonumber	&&  \simeq \frac{\partial}{\partial g} \left< \pmb{\lambda}^{(j)}_{\bar{M}^{(j)}+1,N^{(j)}-M^{(j)}} \right|  H \left| \pmb{\lambda}^{(j)}_{\bar{M}^{(j)}+1,N^{(j)}-M^{(j)}} \right> .  \\\end{eqnarray}
Summing over all the cells, this gives
\begin{eqnarray}
\nonumber	&& C_{{\rm c},>\Lambda} \, :=  \\
\nonumber	&& 2g^2 \frac{\partial}{\partial g} \left( \frac{1}{L} \sum_{j=1}^m \left< \pmb{\lambda}^{(j)}_{\bar{M}^{(j)}+1,N^{(j)}-M^{(j)}} \right|  H \left| \pmb{\lambda}^{(j)}_{\bar{M}^{(j)}+1,N^{(j)}-M^{(j)}} \right> \right) \\
	&& \simeq \, 2g^2 \frac{\partial}{\partial g} \left( \frac{1}{L} \sum_{j=1}^m \left< \pmb{\lambda}_{N^{(j)}} \right|  H \left| \pmb{\lambda}_{N^{(j)}} \right>  \right)  \, = : \, C_{{\rm c}} . 
\end{eqnarray}

Plugging this into Eq.~(\ref{eq:app_CcplusCr_0}) we get the final result
\begin{equation}
w(p) \, \underset{|p| \rightarrow \infty}{\simeq} \, \frac{C_{{\rm c}} + C_{{\rm r}} }{p^4}  ,
\end{equation}
which is our Eq. (4) in the main text.

\section{Calculation of the product $g^2 g^{(2)}(0)$ in the $g\rightarrow \infty$ limit}

In the main text, we use the relation
\begin{equation}
    \label{eq:limg2g2}
    \lim_{g \rightarrow \infty} n^2 g^2 g^{(2)}(0) = 8 \hbar^2/m \, [n e - j^2/(2m)] ,
\end{equation}
where $n = \int \rho(q) dq$ is the particle density, $j = \int  q \rho(q)  dq$ is the momentum density, and $e= \int q^2/(2m) \rho(q) dq$ is the energy density in a state of arbitrary rapidity density $\rho(q)$.
This identity can be derived as follows. We first consider finite $g$. The Hellmann-Feynman theorem, together with thermodynamic Bethe Ansatz calculations (see e.g. Ref.~\cite{kormos2011exact}, or the supplementary methods of Ref.~\cite{malvania2020generalized}), lead to the following formula for $g^{(2)}(0)$, or equivalently for the density of interaction energy $e_{\rm I} := g \partial (E/L)/ \partial g$:
\begin{equation}
    \label{eq:interaction_e}
    e_{\rm I} \, = \, \frac{1}{2}n^2 \, g \, g^{(2)}(0) \, = \,  \int  \left[ q/m - v^{\rm eff}(q)  \right] q\, \rho(q) dq .
\end{equation}
Here $v^{\rm eff}(q)$ is the `effective velocity' defined by the thermodynamic Bethe Ansatz formula
\begin{equation}
    v^{\rm eff}(q) \, = \, \frac{1}{m}\frac{{\rm id}^{\rm dr}(q)}{1^{\rm dr}(q)},
\end{equation}
where ${\rm id}(q) = q$, $1(q) = 1$, and the `dressing' of a function $f(q)$ is defined as
\begin{equation}
    f^{\rm dr}(q) \, = \, f(q) + \int  \varphi(q-q') \frac{f^{\rm dr}(q')}{1^{\rm dr}(q')} \rho(q') dq' .
\end{equation}
Here $\varphi(q) = 2mg / ((mg/\hbar)^2 + q^2)$ is the Lieb-Liniger kernel~\cite{lieb1963exact,lieb_exact_1963}. Expanding at first order in $1/g$, one finds $1^{\rm dr}(q) = 1 + 2\hbar^2 n/(mg) + O(1/g^2)$ and ${\rm id}^{\rm dr}(q) = q + 2\hbar^2 j/(mg) + O(1/g^2)$, so
\begin{equation}
    v^{\rm eff}(q) \, \underset{g \rightarrow \infty}{=} \, \frac{q}{m} - \frac{2\hbar^2}{m^2 g} (q n-j) + O(1/g^2) .
\end{equation}
Inserting this into Eq. (\ref{eq:interaction_e}), one gets the relation (\ref{eq:limg2g2}).

\section{Evolution of the atom density, momentum density and energy density under one-body losses in the hard-core limit}

In the main text we use the fact that, in the hard-core limit, the atom density, momentum density and energy density evolve with time as $n(t) = e^{-G t} n_0$, $j(t) = e^{-G t} j_0$, $e(t) = e^{-G t} e_0$ respectively.

This can be derived using the results of Ref.~\cite{bouchoule_effect_2020} (see also the related  Ref.~\cite{hutsalyuk2020integrability} for the much more difficult case of finite $g$). First, one uses the rapidity distribution to define a generating function for the conserved charges (following Ref.~\cite{bouchoule_effect_2020}),
\begin{equation}
    \label{eq:Qapp}
    Q(z) := \frac{i}{\pi} \int \frac{\rho(q) dq}{z-q} ,
\end{equation}
for $z \in \mathbb{C}$, ${\rm Im} \,z > 0$. $Q(z)$ is analytic for ${\rm Im} \,z > 0$. Moreover, for $q$ real, we have
\begin{equation}
    \lim_{z \rightarrow q} {\rm Re} [ Q(z) ] \, = \, \rho(q).
\end{equation}
Under losses, $Q(z)$ evolves in time. At time $t$, and in terms of the initial rapidity distribution $\rho_0(\lambda)$, it is equal to~\cite{bouchoule_effect_2020}
\begin{equation}
    \label{eq:TG_solutionQ}    	Q(z) \, = \, \frac{ \frac{i  \, e^{-G t} }{\pi \hbar} \int \frac{\rho_0(\lambda) d \lambda}{(z- \lambda)/\hbar + 2 i n_0 (1-e^{- G t})} }{1 -   i  2 (1-e^{-G t})  \int \frac{\rho_0(\lambda) d \lambda}{(z- \lambda)/\hbar + 2 i n_0 (1-e^{- G t})} } ,
\end{equation}
for ${\rm Im} \,z > 0$. 

The atom density $n = \int \rho(q)dq$, the momentum density $j=\int q \rho(q) dq$ and the energy density $e= \int q^2 \rho (q) dq/(2m)$ appear in the asymptotic expansion of Eq.~(\ref{eq:Qapp}) at large $z$:
\begin{equation}
    Q(z) \, \underset{z \rightarrow \infty}{=} \, \frac{i}{\pi} \left( \frac{n}{z} + \frac{j}{z^2} +  \frac{2m e}{z^3} + \dots \right) 
\end{equation}
Expanding Eq.~(\ref{eq:TG_solutionQ}) to order $O(1/z^3)$, one finds
\begin{equation}
     Q(z) \, \underset{z \rightarrow \infty}{=} \, \frac{i}{\pi} \left( \frac{e^{-G t} n_0}{z} + \frac{e^{-G t} j_0}{z^2} +  \frac{2m \, e^{-G t} e_0}{z^3} + \dots \right) ,
\end{equation}
which gives the time-dependence claimed above for the three densities.

\section{Bogoliubov theory in the quasicondensate regime (after Mora and Castin)}

We follow the conventions of Mora and Castin~\cite{mora_extension_2003}. Inserting a phase-amplitude representation of the annihilation operator, $\Psi (z) = \sqrt{n + \delta n} e^{i \theta}$ with $[\delta n(z) , \theta(z')] = i \delta(z-z')$, in the Hamiltonian (2), one finds to second order:
\begin{eqnarray}
 \nonumber   H - \mu N & \simeq & \int \left[ \frac{\hbar^2}{8m n} (\partial_z \delta n)^2  + \frac{g}{2} \delta n^2 + \frac{\hbar^2 n}{2m} (\partial_z \theta)^2  \right] dz .
\end{eqnarray}
This quadratic Hamiltonian allows to grasp quantum fluctuations around the classical profile which solves the Gross-Pitaevski equation, $n = N/L = \mu/g$ where $\mu$ is the chemical potential. One can define a boson annihilation field $B(z) = \frac{1}{2 \sqrt{n}} \delta n(z) + i \sqrt{n} \theta(z)$ such that $[B(z) , B^\dagger(z')] = \delta(z-z')$, and its Fourier modes $B_q = \int e^{-i q z/\hbar} B(z) dz/\sqrt{L}$ with $q\in (2\pi \hbar/L) \mathbb{Z}$. Then the quadratic Hamiltonian becomes, up to constant terms,
\begin{eqnarray}
\nonumber && H - \mu N \, \simeq \\
\nonumber &&   \frac{1}{2} \sum_q \left( \begin{array}{c} B_q \\B_{-q}^\dagger \end{array} \right)^\dagger \left( \begin{array}{cc}
    \frac{q^2}{2m} + \mu &  \mu  \\
   \mu & \frac{q^2}{2m} + \mu 
 \end{array} \right)  \left( \begin{array}{c} B_q \\B_{-q}^\dagger \end{array} \right)  ,
\end{eqnarray}
where we have used $\mu = g n$. Finally, the Hamiltonian $H_q$ is diagonalized by a Bogoliubov transformation
$$
\left( \begin{array}{c}
    B_q \\ B_{-q}^\dagger
\end{array} \right) = \left( \begin{array}{cc}
    \bar{u}_q & \bar{v}^*_{q} \\
    \bar{v}_{-q} & \bar{u}^*_{-q }
\end{array}\right) \left( \begin{array}{c}
    b_q \\ b_{-q}^\dagger
\end{array} \right)
$$
with $|\bar{u}_q|^2 - |\bar{v}_q|^2 = 1$. Here a convenient choice is  $\bar{u}_q = \bar{u}_q^* = \cosh (\theta_q/2)$ and $\bar{v}_q = \bar{v}^*_q = -\sinh (\theta_q/2)$ with
$\tanh \theta_q = \mu/(\mu + \frac{q^2}{2m})$, 
which gives
\begin{eqnarray}
\nonumber && H - \mu N \, \simeq \, \sum_q \varepsilon_q b_q^\dagger b_q \,+\, {\rm const.},
\end{eqnarray}
with a dispersion relation $\varepsilon_q = \sqrt{\frac{q^2}{2m} \left( \frac{q^2}{2m} + \mu \right)}$.

\subsection{Population of Bogoliubov modes and momentum distribution}

Let us consider a state where the population of each Bogoliubov mode is $\alpha_q = \left< b^\dagger_q b_q \right>$. The one-particle density matrix is (see Ref.~\cite{mora_extension_2003}, formula (184)):
\begin{eqnarray}
    \label{eq:opdm}
   && g^{(1)} (z) \, = \, \\
 \nonumber  && \exp \left[-\frac{1}{n} \int \frac{dq}{2\pi \hbar} [ (\bar{u}_q^2 + \bar{v}_q^2) \alpha_q + \bar{v}_q^2 ] (1 - \cos (q z/\hbar) ) \right] .
\end{eqnarray}
Following Lieb~\cite{lieb_exact_1963}, we identify quasiparticle excitations with large rapidities with the large-$q$ Bogoliubov modes. Then we are interested in the case when $n_q$ decays as $2 \pi\hbar  C_{\rm r}/q^4$ at large $q$, where $C_{\rm r}$ is the same constant as in the main text. We note that
\begin{equation}
    [ (\bar{u}_q^2 + \bar{v}_q^2) \alpha_q + \bar{v}_q^2 ] \underset{q \rightarrow \infty}{\simeq} 2\pi \hbar \frac{ C_{\rm r} + C_{\rm c}}{q^4},
\end{equation}
which follows from the fact that $\bar{v}_q^2 = \bar{u}_q^2-1 = m^2 \mu^2/q^4 + O(1/q^6)$, and $m^2 \mu^2 = 2 \pi \hbar C_{\rm c}$ (valid in the quasicondensate regime). In general, $1/k^4$ tails result in a discontinuity of the third derivative of the Fourier transform, according to $\partial_x^3 \left( \int \frac{dk}{2\pi} \frac{e^{i k x}}{k^4 + \epsilon^4} \right)_{|_{x\rightarrow 0^+}} - \partial_x^3 \left( \int \frac{dk}{2\pi} \frac{e^{i k x}}{k^4 + \epsilon^4} \right)_{|_{x\rightarrow 0^-}} = 1$. Thus, the discontinuity of the argument of the exponential in (\ref{eq:opdm}) is
\begin{eqnarray*}
 &&   \partial_z^3 \left( \frac{1}{n} \int \frac{dq}{2\pi \hbar} [ (\bar{u}_q^2 + \bar{v}_q^2) \alpha_q + \bar{v}_q^2 ] (1 - \cos (q z/\hbar) )  \right)_{|_{z \rightarrow 0^+}} \\
 && -  \partial_z^3 \left( \frac{1}{n} \int \frac{dq}{2\pi \hbar} [ (\bar{u}_q^2 + \bar{v}_q^2) \alpha_q + \bar{v}_q^2 ] (1 - \cos (q z/\hbar) )  \right)_{|_{z \rightarrow 0^-}} \\
 && = - 2\pi  \frac{C_{\rm r} + C_{\rm c}}{\hbar^3 n} .
\end{eqnarray*}
Consequently, $g^{(1)}(z)$ also possesses a discontinuity in its third derivative,
\begin{equation}
    \partial_z^3 g^{(1)}_{|_{z \rightarrow 0^+}} -  \partial_z^3 g^{(1)}_{|_{z \rightarrow 0^-}} \, = \, 2\pi  \frac{C_{\rm r} + C_{\rm c}}{\hbar^3 \rho_0} .
\end{equation}
Taking the Fourier transform, one finds that the momentum distribution has a tail with coefficient $C_{\rm r} + C_{\rm c}$, as claimed in the main text:
\begin{eqnarray}
 \nonumber   w(p) &=& \frac{n}{2 \pi \hbar} \int_0^L e^{i p z/\hbar} g^{(1)}(z) dz \\
    & \underset{p \rightarrow \infty}{\simeq} & (C_{\rm r} + C_{\rm c})/p^4 .
\end{eqnarray}

\subsection{The effect of losses on Bogoliubov modes}

The effect of losses in the quasicondensate regime has been investigated in Refs.~\cite{grisins_degenerate_2016,johnson_long-lived_2017,schemmer_monte_2017,bouchoule_cooling_2018}. For the convenience of the reader, we recall the results that are useful for this Letter. 

In terms of the Fourier modes of the phase and density fluctuation fields, $\theta_q=(1/\sqrt{L})\int dz \theta(z)e^{-iqz/\hbar}$ and $\delta n_q=(1/\sqrt{L})\int dz \delta n(z)e^{-iqz/\hbar}$, the population $\alpha_q$ of the Bogoliubov mode $q$ reads
\begin{equation}
    \alpha_q = 
    \frac{f_q}{4 n}  \langle\delta n_{-q}\delta n_q \rangle + \frac{n}{f_q} \langle\theta_{-q}\theta_q \rangle -\frac{1}{2} ,
    \label{eq:Hqdelantheta}
\end{equation}
where $f_q=\sqrt{(q^2/(2m) + 2gn)/(q^2/(2m))}$. 

Under losses, the density $n$ and the coefficient $f_q$ become time-dependent, as well as the phase and density fluctuations $\langle \delta n_{-q}\delta n_q\rangle$ and 
$\langle \theta_{-q}\theta_q\rangle$. One finds
\begin{eqnarray}
    \frac{d \alpha_q}{dt}& =&  \frac{f_q}{4 n} \frac{ d \langle\delta n_{-q}\delta n_q \rangle }{dt} + \frac{n}{f_q} \frac{ d \langle\theta_{-q}\theta_q \rangle }{dt} \label{eq:dalphaqdtcomplete}
    \\
    && + \frac{1}{f_q/n} \frac{d (f_q/n)}{dt} \left[ \frac{f_q}{4 n}  \langle\delta n_{-q}\delta n_q \rangle - \frac{n}{f_q} \langle\theta_{-q}\theta_q \rangle \right] . \nonumber
\end{eqnarray}
We are assuming slow losses. Then, to compute  
$d\alpha_q/dt$, which is a slowly varying quantity, 
one can average over a time $2\pi/\varepsilon_q$. 
This time-average ensures equipartition of energy between 
the two conjuagte variables $\delta n_q$ and $\theta_{-q}$.
Consequently, the second line in the equation vanishes, and we have
\begin{equation}
\label{eq:dalphadtSM}
    \frac{d \alpha_q}{dt} =  \frac{f_q}{4 n} \frac{ d \langle\delta n_{-q}\delta n_q \rangle }{dt} + \frac{n}{f_q} \frac{ d \langle\theta_{-q}\theta_q \rangle }{dt} ,
\end{equation}
which is the equation used in the main text. 
Note that the fact that $d\alpha_q/dt$ is not affected by the 
slow time evolution of $n$ and $f_q$ ({\it i.e.} the vanishing of the second line of Eq.~\eqref{eq:dalphaqdtcomplete}) 
can also be interpreted 
as the 
result of adiabatic following of the eigenstates of 
$H_q=\varepsilon_q (b_q^+b_q +1/2)$.
We now recall the effect of losses 
on density and phase fluctuations, analyzed in Refs.~\cite{bouchoule_cooling_2018}.

\subsubsection{Effet of losses on density fluctuations}

The goal of this section is to derive the formula for the evolution of the density fluctuations,
\begin{eqnarray}
    \label{eq:ddeltaN2dt}
  &&  \frac{d\langle \delta n (z) \delta n(z') \rangle}{dt} = 
    K^2 G n^K   \delta (z-z') \\
\nonumber     &&  \qquad \qquad \qquad \qquad - 2K^2 G n^{K-1} \langle \delta n(z) \delta n(z') \rangle ,
\end{eqnarray}
which is used in the main text.

To do this, we consider a cell of length $\ell$, much smaller than the typical length scale of variation of the phase $\theta$, but large enough so that it contains a number of atoms $N \gg 1$.
We note $\bar{N} = n \ell$ the atom number corresponding to the mean atomic density $n$ in the gas. 
We are interested in the effect of losses during a time interval $\Delta t$ satisfying $\bar{N}^{-K} \ll \gamma \Delta t \ll \bar{N}^{1-K}$ where $\gamma :=G/\ell^{K-1}$ is the loss rate in the cell. This ensures that the number of lost atoms is much larger than one, but much smaller than $\bar{N}$.

We consider an initial state with an atom number distribution $P_0(N)$. Here fluctuations can be either of statistical or of quantum nature. Let ${\cal P}_0(M)$ the probability to have $M$ loss events until time $\Delta t$. One has
\begin{equation}
    {\cal P}_0(M)=\sum_N P_0(N)P(M|N),
\end{equation}
where $P(M|N)$ is the probability to have $M$ loss events conditioned to an initial number of atoms $N$. Under the assumption $\gamma \Delta t \ll \bar{N}^{1-K}$, this is well approximated by a Poisson distribution~\cite{schemmer_monte_2017}
\begin{equation}
    P(M|N)=\frac{1}{M!}e^{-\gamma \Delta t \, N^K}\left ( \gamma \Delta t \, N^K\right )^M.
\end{equation}
Furthermore, for $ \gamma \Delta t \gg \bar{N}^{-K} $, the Poissonian becomes a Gaussian,
\begin{equation}
	\label{eq:PMNgauss}
    P(M|N)\simeq \frac{e^{-(M- N^K  \gamma \Delta t )^2}}{\sqrt{2\pi}\sigma} .
\end{equation}
The variance can be approximated by its value for $N= \bar{N}$, which is
\begin{equation}
    \sigma =\sqrt{\gamma \Delta t \, \bar{N}^K}.
\end{equation}
The probability to have $N$ atoms in the cell at time $\Delta t$ is then
\begin{eqnarray}
\nonumber    P(N) &=& \sum_M P_0(N+KM)P(M|N+KM) \\
    	& \simeq & \int dM P_0(N+KM)P(M|N+KM) , \quad 
    \label{eq:PN}
\end{eqnarray}
where we have used the fact that both $N$ and $M$ are typically large to replace the sum by an integral. 

We are now ready to compute the atom number fluctuations at time $\Delta t$. For this we introduce
$\delta N(0) =N -\bar{N}$ at time $0$, and  
$\delta N( \Delta t)=N( \Delta t )-\bar{N}( \Delta t )$ at time $\Delta t$,  where 
$\bar{N}( \Delta t )=\bar{N}-K  \gamma \Delta t \, \bar{N}^K$ is the atom number 
corresponding to the gas mean density after $\Delta t$.
Using \eqref{eq:PN}, one gets
\begin{equation*}
\begin{array}{ll}
    \langle \delta N(\Delta t)^2\rangle =
    \int dN \int& dM (N-\bar{N}(\Delta t))^2 \\
    &P_0(N+KM)P(M|N+KM)
    \end{array}
\end{equation*}
With the change of variable $\tilde{N}=N+KM$, this becomes
\begin{eqnarray}
   && \langle \delta N(\Delta t)^2\rangle \, = \\
\nonumber  &&  \int d\tilde{N} P_0(\tilde{N})\int dM (\tilde{N}-KM-\bar{N}(\Delta t))^2 P(M|\tilde{N}).
\end{eqnarray}
Then the Gaussian approximation of $P(M | \tilde{N})$ (Eq.~\eqref{eq:PMNgauss}) gives
\begin{eqnarray}
   && \langle \delta N( \Delta t )^2\rangle = 
    K^2 \gamma \Delta t \, \bar{N}^K \\
\nonumber & &  \qquad +  \int d\tilde{N} P_0(\tilde{N}) (\tilde{N}-K \gamma \Delta t  \, \tilde{N}^K -\bar{N} +K \gamma \Delta t \, \bar{N}^K)^2 .
\end{eqnarray}
Using the fact that the atom number fluctuations around $\bar{N}$ are small, one can expand to lowest order in 
$\delta N(0)=\tilde{N}-\bar{N}$. Then the expression inside the parenthesis becomes $(1-K^2\gamma \Delta t \bar{N}^{K-1}) \, \delta N(0) $; the square of that expression is $(1-2 K^2 \gamma \Delta t \bar{N}^{K-1}) \, \delta N(0)^2$ at first order in $\gamma \Delta t \, \bar{N}^{K-1}$. Thus we obtain
\begin{equation}
    \langle \delta N(\Delta t)^2\rangle =
    K^2 \gamma \Delta t \, \bar{N}^K + \left ( 1-2K^2 \gamma \Delta t \, \bar{N}^{K-1}\right ) \langle \delta N^2\rangle
.
\end{equation}
This lead to the differential form
\begin{equation}
    \frac{d\langle \delta N^2\rangle}{dt} =
    K^2 G n^K\ell - 2K^2 G n^{K-1} \langle \delta N^2\rangle ,
    \label{eq:deltaNDeltat1cell}
\end{equation}
where we have used $\gamma \bar{N}^{K-1} = G n^{K-1}$.

Let us now consider two differents cells located around $z_\alpha$ and $z_\beta$. 
For given atom numbers $N_\alpha$ and $N_\beta$ in the cell located in $z_\alpha$ and $z_\beta$ respectively, the fluctuations of the number of loss events in both cells are not correlated. 
Then similar calculations as above give
\begin{equation}
    \frac{d\langle \delta N_\alpha \delta N_\beta \rangle}{dt} =
    -2K^2 \gamma \Delta t \, \bar{N}^{K-1} \langle \delta N_\alpha \delta N_\beta \rangle .
\label{eq:deltaNDeltat2cell}
\end{equation}

Eq.\eqref{eq:deltaNDeltat1cell} and \eqref{eq:deltaNDeltat2cell} imply  that the evolution of the fluctuations of the density field $\delta n (z) \simeq \delta N/\ell$ (for a cell around at position $z$) is given by Eq.~(\ref{eq:ddeltaN2dt}) as claimed.

\subsubsection{Effect of losses on phase fluctuations}

Although losses do not depend on the phase variable, losses do have an impact on the phase fluctuations 
$\langle \theta(z)^2 \rangle$. This is due to the broadening of the phase as one gains knowledge on the atom number $N$, its conjugate variable. This ensures the preservation of quantum uncertainty relations. Losses increase our knowledge of $N$ because if one records the losses, then one gains knowledge on $N$ [This effect can be exploited in a feedback scheme to cool down the Bogoliubov modes~\cite{schemmer_monte_2017}]. The quantitative evaluation of this effect is done in Ref.~\cite{bouchoule_cooling_2018}, and the result reads:
\begin{equation}
    \frac{d\langle \theta(z) \theta(z') \rangle}{dt}=   \frac{1}{4} K^2G n^{K-2} \delta (z-z').
\label{eq:dtheta2dt}
\end{equation}
This is the equation used in the main text. [We point out that Eq.~\eqref{eq:dtheta2dt}, as well as Eq.~\eqref{eq:ddeltaN2dt}, can also be derived from stochastic equations, see Ref.~\cite{bouchoule_cooling_2018}.]

\subsubsection{Evolution of the population of the Bogoliubov modes}
Taking the Fourier transform of Eq.~\eqref{eq:ddeltaN2dt} and
Eq.~\eqref{eq:dtheta2dt} and injecting into 
Eq.~\eqref{eq:dalphadtSM} we find 
\begin{equation}
    d  \alpha_q/dt = K^2Gn^{K-1} \left ( -\alpha_q -1/2 +1/4(f_q+f_q^{-1})\right ). 
    \label{eq:dalphadtvsfq}
\end{equation}
This equation, together with the equation
$n=n_0e^{-Gt}$, allows to compute $\alpha_q(t)$.
This equation is valid for any value of $q$.

\subsection{Evolution of the momentum distribution}
\begin{figure}[h]
    \centering
    \includegraphics[width=0.48\textwidth]{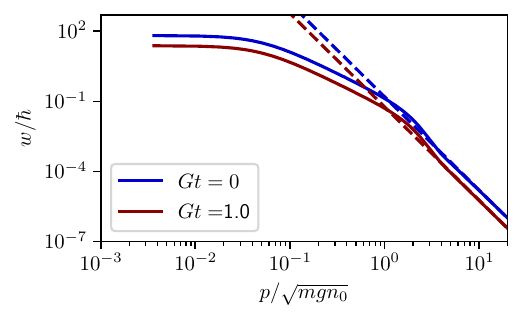}
    \caption{ Momentum distribution of a quasicondensate submitted to one-body losses of rate $G$. The initial state is a thermal state at a linear density $n_0=10\sqrt{mgn_0}/\hbar$ and at a temperature $T=gn_0$. Its momentum distribution is shown as the blue solid line. The dashed blue line is $C_{c,0}/p^4$, where $C_{c,0}=(mn_0g)^2/(2\pi)$ is the initial
    contact density ($g^{(2)}(0)\simeq 1$ in the quasicondensate regime). The red solid line is the momentum distribution after a time $t=1/G$. 
    The dashed red line is $C(t)/p^4$, where $C(t)=e^{Gt}(mn_0g)^2/(2\pi)$.}
    \label{fig:wpBogo}
\end{figure}

We performed numerical calculations for one-body losses ($K=1$), starting from a thermal state with linear density $n_0$ and temperature $T$.
We use Eq.~(\ref{eq:dalphadtvsfq}), injecting $n(t)=n_0e{-Gt}$, to compute
$\alpha_q(t)$ for all $q$. 
We then compute the first order correlation function using
Eq.\eqref{eq:opdm}. We finally 
take its Fourier transform to extract the 
momentum distribution $w(p)$.
Fig.\ref{fig:wpBogo} shows resulting momentum 
distributions, in log-log scale, at time $t=0$ and at time 
$t=1/\Gamma$. 
We see that, for those parameters, 
the $1/p^4$ behavior appears for momenta larger than 
$\simeq 3 \sqrt{mgn_0}$. The amplitude of the tails 
is in agreement with the analytic prediction 
$C(t)=e^{Gt}(mn(t)g)^2/(2\pi)$.

\subsection{Solution of the differential equation (13) for losses in the quasicondensate regime}

We use the dimensionless variable $\tau = K n_0^{K-1} G t$, where $n_0$ is the atom density at $t=0$. In the quasicondensate regime, we have $g^{(K)}(0) = 1$, so the atom density $n(\tau)$ evolves according to
\begin{equation}
    \label{eq:dnn0}
    \frac{d (n/n_0)}{ d\tau} = - (n/n_0)^{K} .
\end{equation}
The differential equation (13) in the main text is
\begin{equation}
   \frac{d C_{\rm r}}{d\tau} \, = \, - K (n/n_0)^{K-1}  C_{\rm r} + K C_{{\rm c},0} \, (n/n_0)^{K+1}  ,
\end{equation}
with $C_{{\rm c},0} =  m^2 g^2 n_0^2/(2\pi \hbar)$. Using Eq.~(\ref{eq:dnn0}) one can easily check that the solutions of that differential equation are (for $K \neq 2$)
\begin{equation}
    C_{\rm r}(\tau) \, = \, \frac{K \, C_{{\rm c},0} }{K-2} (n/n_0)^2 + A \, (n/n_0)^K,
\end{equation}
for any constant $A$. The constant $A$ is then fixed in terms of the initial condition $C_{\rm r}(t=0) = 0$ (this is the initial condition assumed in the main text). This gives (for $K \neq 2$):
\begin{equation}
    C_{\rm r}(\tau) \, = \,  \frac{K \, C_{{\rm c},0}}{K-2} (n/n_0)^2  \left[1 - (n/n_0)^{K-2} \right] .
\end{equation}
If $K=2$, then we have instead
\begin{equation}
(K=2) \quad C_{\rm r}(\tau) \, = \, - 2  C_{{\rm c},0} (n/n_0)^2 \log (n/n_0) .
\end{equation}
Recall that $C_{{\rm c}}(\tau) = m^2 g^2 n(\tau)^2/(2\pi \hbar)$. Then we get
\begin{equation}
    \label{eq:CrCc_app}
    \frac{C_{\rm r}(\tau)}{C_{\rm c}(\tau)} \, = \,  \left\{ \begin{array}{ccl}
            K/(K-2) \, \left[1 - (n/n_0)^{K-2} \right] & {\rm if} & K \neq 2 , \\
            - 2 \log (n/n_0) & {\rm if} & K=2 .
        \end{array} \right.
\end{equation}
Finally, we note that the solution of Eq.~(\ref{eq:dnn0}) is
\begin{equation}
    \label{eq:ntau}
   \frac{ n(\tau)}{n_0} \, = \, \left\{ \begin{array}{ccl}
        \left[ 1 +  (K-1) \tau \right]^{ 1/(1-K)}  & {\rm if}& K > 1 , \\
        e^{-\tau} & {\rm if} & K=1 .
    \end{array} \right.
\end{equation}
Eqs.~(\ref{eq:CrCc_app}) and (\ref{eq:ntau}) give the large $\tau$ behavior reported in Eq.~(14) in the main text.

\section{Generalization to non-uniform gases}
In most experimental situations, gases are confined into a slowly-varying
longitudinal potential, often of quadratic form. 
The confinement is however usually weak enough to ensure the validity 
of the Generalized Hydrodynamics approach~\cite{bertini2016transport,castro2016emergent} (which corresponds, in the case of stationary states, to the well known Local Density Approximation). 
The rapidity distribution then becomes a two dimensional function $\rho(q,z)$,
where, for a given $z$, $\rho(q,z)$ is the local rapidity distribution.
The coefficient $C_r=\lim_{q \rightarrow \infty} q^4 \rho(q)$ becomes 
$z$-dependent and we note it $C_r(z)$.
Moreover we introduce the extensive quantity $W(p)=\int dz~w(p,z)$, where 
$w(p,z)$ is the local momentum distribution, and 
${\cal C}=\lim_{p\rightarrow \infty} p^4 W(p)$. $W(p)$ is normalized to 
$\int dp~W(p)=N$ where $N$ is the total atom number. 
Eq.~(4) of the main text then becomes
\begin{equation}
    {\cal C}=\int dz \left ( C_{\rm c}(z) +C_r(z)\right )
\end{equation}
where $C_{\rm c} (z)=m^2g^2n(z)^2g^{(2)}(0,z)/(2\pi\hbar)$ is the 
local contact density.
Here  $g^{(2)}(0,z)=\langle \psi^+(z)\psi^+(z)\psi(z)\psi(z)\rangle$  is the zero-distance two-body correlation function, computed at 
position $z$.
For a given $z$, $C_{\rm c}(z)$ is a functional of $\rho(p,z)$, see Eq.~(\ref{eq:interaction_e}). Thus 
$C$ can be computed once the function $\rho(p,z)$ is known.

As losses occur, $\rho(p,z)$ is locally modified by losses. The system  is
then, in general, brought to a non-stationary solution of the Generalized Hydrodynamics equations and one should compute the time-evolution of
$\rho(p,z)$ using Eq.~(16) of Ref.~\cite{bouchoule_effect_2020}.  


\end{document}